\documentstyle[aps,prb,twocolumn,graphicx]{revtex}

\begin{document}
\draft
\title{Lattice dependence of saturated ferromagnetism in the Hubbard model}

\author{Thoralf Hanisch, G\"otz S.~Uhrig and Erwin M\"uller-Hartmann}
\address{Institut f\"ur Theoretische Physik, Universit\"at zu K\"oln,
  D-50937 K\"oln}

\date{\today}

\maketitle

\begin{abstract}
We investigate the instability of the saturated ferromagnetic ground state
(Nagaoka state) in the Hubbard model on various lattices in dimensions
$d=2$ and $d=3$. A variational resolvent approach is developed 
for the Nagaoka instability both for $U=\infty$ and for $U<\infty$
which can easily be evaluated in the thermodynamic limit on all common 
lattices. Our results significantly improve former variational bounds for 
a possible Nagaoka regime in the ground state phase diagram of the 
Hubbard model. We show  that a pronounced particle-hole asymmetry in the 
density of states and a diverging density of states at the lower band 
edge are the most important features in order to stabilize Nagaoka 
ferromagnetism, particularly in the low density limit.

\end{abstract}

\section{Introduction} \label{sect:intro}

It is by now an often repeated fact that the  so-called 
(single-band) Hubbard model
was originally introduced to explain ferromagnetism 
\cite{gutzw63,hubba63,kanam63}. In what followed, however, it turned
out to be rather a generic model for antiferromagnetism.
Ferromagnetism seemed to require additional ingredients, for instance
the existence of degenerate bands which favor ferromagnetism based on
Hund's rule or in the insulating case certain additional ferromagnetic 
couplings and/or correlated hopping terms. Both scenarios were proven
 rigorously 
in recent years (see  \cite{mielk93b} and references therein for the former
 and \cite{kolla96} and references therein for the latter).

The  Hubbard model and its possible ferromagnetic ground 
state are of renewed interest \cite{fazek97,vollh97}.
 There are many works in the
field based on quasi one dimensional  ($d=1$) systems triggered by the
prediction of ferromagnetism in double minima systems at
low particle density \cite{mulle95} and by the numerous possibilities
of analytical and numerical calculations 
\cite{penc96a,pieri96,daul96,sreer97} in $d=1$.
Exact calculations are possible
in infinite dimensions ($d=\infty$) \cite{uhrig96a,ulmke96}. For
intermediate dimensions ($1< d < \infty$) numerical and approximate methods
are employed \cite{hlubi97,herrm97a,herrm97b}.

An important mile stone in the research of ferromagnetism in Hubbard
models is the work of Nagaoka \cite{nagao65,tasak89}. It showed that at
infinite local repulsion a single electron above half-filling favors
the saturated ferromagnetic ground state (henceforth: Nagaoka state)
if the underlying lattice has loops which allow interference.
For bipartite lattices particle-hole symmetry extends these
results to hole doping. This result reveals the beauty and the difficulty
of the question for which lattices and for which fillings the Nagaoka
state is the ground state. At $U=\infty, T=0$ there is only the hopping 
left as a global energy scale. Thus there is no expansion parameter,
 no adiabatic limit, and
no competition of energy scales. The issue is solely a question of
the lattice structure, i.e.\ the possible paths on the lattice, and of
the filling.

Unfortunately, there are no extensions of Nagaoka's result to 
macroscopic dopings. Only non-macroscopic numbers of holes could be treated
\cite{barbi90,tian91}. Therefore, we choose another route in the present
work and investigate the stability of the Nagaoka state towards a single
spin flip. If such a flip lowers the energy then the Nagaoka state is not
the ground state. Otherwise it is locally stable. The drawback that we
treat only local stability in this way is not very serious. There is no
indication that the transition away from saturation should not be 
continuous at $T=0$, see e.g. \cite{takah69}. 

A more serious drawback is the fact that even the single spin flip
is too difficult a problem to be solved completely on finite dimensional
lattices. In the limit of infinite dimensional lattices, however,
it was solved \cite{uhrig96a}. Thereby it was shown that 
relatively simple variational ansatzes provide already a qualitative
insight in the tendency of a certain lattice to have a Nagaoka state
as ground state. Wurth {\em et al}. showed that only
an extremely sophisticated variational ansatz \cite{wurth96}
yields a further reduction
of the region of possible Nagaoka state stability 
in comparison to simpler ansatzes \cite{hanis93}.

It is the aim of the present paper to extend previous work 
on variational ansatzes decisively \cite{hanis95}, both in the
completeness of the ansatzes and in the types of the lattices considered.
So far, variational ansatzes considered a finite vicinity of the flipped
spin and treated a finite number of parameters leading to matrix
eigenvalue problems. Here we will show that a resolvent approach is
capable to deal implicitly with an \textit{infinite} number of
variational parameters. No explicit knowledge of the variational
wave function is required. A similar approach was used recently 
by Okabe\cite{okabe97} for the square lattice and the 
simple cubic lattice, too. In his work,
however, the reduction of the resolvent to simple integrals over the 
density of states (DOS), 
which we succeeded to achieve in most cases, is lacking.

We will present elegant simple expressions for the 
Nagaoka instability line
$U_{\rm cr}(n)$ which apply to most common lattices. These results
make it possible for everyone to check easily whether or not one can
expect a ferromagnetic ground state for a given lattice. 
We will show that two main features favor the occurrence of
a saturated ferromagnetic ground state
\begin{enumerate}
\item
a highly asymmetric density of states with large values at the
lower band edge (after particle-hole transformation).
\item
non-bipartiteness of the lattice, i.e.\ frustration due to loops of 
three sites.
\end{enumerate}
Of course, the two points are intimately related. 

The setup of our article is as follows. In the rest of the Introduction
we will present certain variational ansatzes used so far to investigate
the Nagaoka state stability. In the subsequent section \ref{sect:resapp}
we develop the resolvent approach which yields simple formulae
for the stability lines on homogeneous, isotropic lattices 
with nearest neighbor hopping. In sect.\ \ref{sect:resvarlat} 
we present our
results for various lattices in dimensions $d=2$ and $d=3$,
namely the square, the simple cubic, the \textit{bcc}, the honeycomb, 
the triangular, the kagome, and the \textit{fcc} (\textit{hcp}) lattice. For
the $t$-$t'$ Hubbard model on the square lattice a perturbative approach
for small $|t|$ is employed as well. Sect.\ 
\ref{sect:conclus} contains
a summary and a final discussion of the lattice dependence of saturated
ferromagnetism in the Hubbard model.
The appendices contain technical details in the derivation for the various 
lattices.

\subsection{Preliminary approaches}

We consider the conventional single band Hubbard model
\begin{equation} \label{def:model}
H = -t \sum_{\langle \underline{i}, \underline{j}\rangle \sigma}
a_{\underline{i} \sigma}^{+} a_{\underline{j} \sigma} + U 
\sum_{\underline{i}} a_{\underline{i} \uparrow}^{+}
a_{\underline{i} \uparrow} a_{\underline{i} \downarrow}^{+}
a_{\underline{i} \downarrow}
\end{equation}
and calculate the spin flip energy
\begin{equation} \label{def:esf}
\Delta e = \langle \Psi | H - E_{\cal N} | \Psi \rangle / \langle \Psi |
\Psi \rangle
\end{equation}
where $E_{\cal N}$ is the energy of the Nagaoka state and $|\Psi \rangle$ is
a variational wave function.  Whenever $\Delta e < 0$
the Nagaoka state is definitely not the ground state due to the variational
nature of our approach. At $U = \infty$, the zero of $\Delta e_{\infty}
(\delta) := \Delta e (U=\infty, \delta)$ gives the critical hole density 
$\delta_{\rm cr}$ above which the Nagaoka state is unstable. For finite $U$, 
$\Delta e (U, \delta) \doteq 0$ leads to the Nagaoka instability 
line $U_{\rm cr} (\delta)$
which separates a region of guaranteed instability of the Nagaoka state 
($U<U_{\rm cr}(\delta)$) in the phase diagram of the Hubbard model
from a region of possible stability of the Nagaoka state 
($U>U_{\rm cr}(\delta)$).
In the phase diagrams displayed in this paper we will always represent the
on-site repulsion $U$ in terms of $U_{\rm red} = U/(U+U_{\rm BR})$ where
$U_{\rm BR} = -16 \epsilon^{0}$ denotes the Brinkman-Rice critical 
coupling\cite{brink70}. $\epsilon^{0}$ is
the energy per particle of the saturated ferromagnetic state for the 
quarter-filled band and depends on the underlying lattice.
This representation is chosen to render comparisons between different
lattices possible.

A common starting point\cite{shast90a,mulle91,mulle93b,hanis93,hanis95}
 is defined by the ansatz
\begin{eqnarray} \nonumber
| \Psi \rangle &:=& |\Lambda|^{-1/2} \sum_{\underline{i}}
\exp( i \underline{k}_{\rm b} \underline{i}) \,
\big[a_{\underline{i} \uparrow} \big(a_{\underline{i} \uparrow}^{+} +
f \cdot \sum_{\langle \underline{i}, \underline{j}\rangle} 
a_{\underline{j} \uparrow}^{+}\big)
\\ \label{ansatz:gw}
&+&
g \cdot a_{\underline{i} \uparrow}^{+} a_{\underline{i} \uparrow}\big] 
a_{\underline{i} \downarrow}^{+} | {\cal N}' \rangle \ .
\end{eqnarray}
For $f=0$  this is the Gutzwiller single spin flip wave function (Gw). The
parameter $g$ controls the probability of double occupancy.
The system size is denoted by $|\Lambda|$. 
We use the operators $a \, (a^{+})$ for site diagonal fermion 
annihilation (creation)
and $c \, (c^{+})$ for momentum diagonal fermion annihilation (creation). 
Furthermore, we use $n$ for the particle density, $\delta = 1-n$ for
the doping per site, $z$ for the coordination number, and 
$e_1=E_{\cal N}/|\Lambda|$
 for the expectation value of the kinetic energy.
The ket $| {\cal N}' \rangle 
= c_{\underline{k}_{\rm F} \uparrow} | {\cal N} \rangle$ is the fully 
polarized Fermi sea of $\uparrow$-electrons from which one 
${\rm e}_{\uparrow}^{-}$ at the Fermi level $\varepsilon_{\rm F}$ 
is removed. 
The energy balance of (\ref{ansatz:gw}) with $f=0$
reads at infinite $U$ ($g=0$) (see (\cite{shast90a,hanis93})
\begin{equation}
\label{skaband}
\Delta e = -e_1/\delta -\varepsilon_{\rm F} + 
\varepsilon_{\underline{k}}\delta(1-(e_1/\delta z t)^2)
\end{equation}
where $\varepsilon_{\underline{k}}$ is the dispersion.
The maximum energy lowering is obviously obtained for $\underline{k}$
 belonging 
to the lower band egde $\varepsilon_{\rm b}$, i.e.\ here 
$\underline{k}_{\rm b}=\underline{0}$.

For finite $f$
 majority spin hopping processes from the position of the flipped spin
to nearest neighbor sites are taken into account. This ansatz will be
denoted NN. The amplitudes of these hopping
processes are assumed to reflect the lattice symmetry.
Basile and Elser investigated an ansatz similar to NN which
includes $\uparrow$-hopping processes from the position of the 
$\downarrow$-electron to \textit{all} other lattice sites \cite{basil90}. 
Since the number of variational parameters increases with the lattice size 
they only studied a finite square lattice. The resolvent method developed 
in sect.\ \ref{sect:resapp} allows us to investigate a variational ansatz 
equivalent to the full Basile-Elser wave function \textit{in the 
thermodynamic limit} on all common lattices. We also derive 
improved variational criteria for the Nagaoka instability at $U<\infty$ by
 extending the Hilbert subspace further.

\section{Resolvent approach} \label{sect:resapp}
Generally, a resolvent is an operator-valued expression
of the type 
a\begin{equation}
\label{resallg}
R(\omega) = 1/(\omega-(H-E_{\cal N}))
\end{equation}
 where $H-E_{\cal N}$ is the
Hamilton operator with respect to the ground state
energy $E_{\cal N}$ (here: the Nagaoka state energy).
 From (\ref{resallg}) it is clear that the existence of
any state at $\omega=\omega_0$ implies a pole or at least
a singularity in the resolvent. For this reason, we will
investigate in the following the resolvent $R$ applied to
$c^+_{\underline{k}_{\rm b}\downarrow}|{\cal N}'\rangle$
and compare $\omega_0$ to $\varepsilon_{\rm F}$.

It is not possible to compute $R$ for the whole
Hilbert space except under simplifying conditions like
infinite coordination number \cite{uhrig96c}. Hence we
will restrict the inversion to certain subspaces which still allow an
analytical treatment. The results obtained in this way
for the lower band edge are variational.
This means that excitation energies found 
 are upper bounds to the true ones
and that specific interaction values $U$ come
out too small compared to those of the full solution.

\subsection{Case $U=\infty$: Ansatz RES0}
For infinite on-site repulsion no double occupancy is allowed.
Thus at the site of the $\downarrow$-e$^-$ no $\uparrow$-e$^-$
is allowed. We investigate therefore the variational subspace
spanned by 
$a_{\underline{i}\uparrow}^{\phantom{+}} a^+_{\underline{j}\uparrow}  
a^+_{\underline{i}\downarrow}|{\cal N}'\rangle$ with 
arbitrary $\underline{i}$ and $\underline{j}$. 
We define 
\begin{mathletters}
\label{varia}
\begin{equation}
| \Phi_{\underline{k}} \rangle
 := {\cal A}_{\underline{k}} | {\cal N}' \rangle 
\end{equation}
\begin{equation}
{\cal A}_{\underline{k}}  := 
|\Lambda|^{-1/2}\sum\limits_{\underline{i}}
 \exp(i(\underline{k}_{\rm b}- \underline{k})\underline{i})
a_{\underline{i}\uparrow}^{\phantom{+}} 
c_{\underline{k}\uparrow}^{+}
a_{\underline{i}\downarrow}^{+}  \ ,
\end{equation}
\end{mathletters}
where the admissible values of $\underline{k}$ are
outside the Fermi sphere (FS), but inside the Brillouin zone (BZ),
i.e.\ $\underline{k} \in {\rm BZ} \setminus {\rm FS}$.
The Hamiltonian does not mix states (\ref{varia}) for
{\em different} total momenta $\underline{k}_{\rm b}$. 
States (\ref{varia}) for
{\em different} total momenta $\underline{k}_{\rm b}$ are orthogonal.
Ansatz (\ref{varia}) contains in particular the NN ansatz (\ref{ansatz:gw})
and of course the simple Gutzwiller ansatz. It comprises
$\uparrow$-hopping processes of arbitrary distance, 
i.e.\ it is the thermodynamic extension
of the ansatz investigated previously by Basile and Elser \cite{basil90}.

For the computation of the resolvent $R(\omega)$ one can use
the Mori/Zwanzig projection formalism 
(see e.g.\ appendix C in \cite{fulde93}) with the scalar product
$({\cal A} |{\cal B}) := \langle {\cal N}' | 
[{\cal A}^+,{\cal B}]_+ | {\cal N}' \rangle$ for the operators 
${\cal A}$ and ${\cal B}$.
The resolvent (\ref{resallg}) then becomes
\begin{eqnarray}\nonumber
{\bf R}_{\underline{k}_{1}, \underline{k}_{2}} (\omega) &=& 
\langle \Phi_{\underline{k}_{1}}| R(\omega) 
|\Phi_{\underline{k}_{2}} \rangle
\\ \label{resmori}
&=&
({\mathcal{A}}_{\underline{k}_{1}} |
(\omega - {\cal L})^{-1} {\mathcal{A}}_{\underline{k}_{2}})\ .
\end{eqnarray}
Here the Liouville operator ${\cal L}$ is used which is defined as
${\cal LA}:=[H,{\cal A}]$ for all operators ${\cal A}$ \cite{fulde93}.
The resolvent can be expressed in matrix notation \cite{fulde93} by
\begin{equation}
\label{resolv}
{\bf R}(\omega) = {\bf P} \, (\omega {\bf P} - 
{\bf L} - {\bf M}(\omega))^{-1}
\, {\bf P}
\end{equation} 
with the norm matrix ${\bf P}$ and the frequency matrix ${\bf L}$
\begin{mathletters}
\begin{equation}
{\bf P}_{\underline{k}_{1}, \underline{k}_{2}} := 
\langle \Phi_{\underline{k}_{1}} | \Phi_{\underline{k}_{2}} \rangle
\end{equation}
\begin{equation}
{\bf L}_{\underline{k}_{1}, \underline{k}_{2}} := 
\langle \Phi_{\underline{k}_{1}} | H -
E_{\cal N} | \Phi_{\underline{k}_{2}} \rangle \ .
\end{equation}
\end{mathletters}
The frequency matrix ${\bf L}$ encodes the effect of $H$ in the subspace
considered. The deviation of  ${\bf P}$ from unity accounts for the
non-orthonormality of the basis. The so-called memory matrix 
${\bf M}(\omega)$ describes the effect of all processes which
imply excursions outside the subspace considered. If the ground
state is known exactly (which holds in the present case) the
approximation ${\bf M}(\omega)={\bf 0}$ is variational in nature
for the lower band edge.

It is the aim of the subsequent calculation to obtain a simple
condition for the singularity of $(\omega {\bf P} - {\bf L})$.
This singularity then signals that $\omega$ corresponds to 
an eigen energy. To this end, we first need the matrix elements
\begin{mathletters}
\label{xelem}
\begin{equation}
\label{xelema}
{\bf P}_{\underline{k}_{1}, \underline{k}_{2}} = 
n \, \delta_{\underline{k}_{1}, \underline{k}_{2}} + |\Lambda|^{-1}
\end{equation}
\begin{equation}
\label{xelemb}
{\bf L}_{\underline{k}_{1}, \underline{k}_{2} \uparrow} = 
\delta_{\underline{k}_{1}, \underline{k}_{2}} (n \cdot 
\varepsilon_{\underline{k}_{2}} - e_1)
\end{equation}
\begin{equation}
\label{xelemc}
{\bf L}_{\underline{k}_{1}, \underline{k}_{2} \downarrow}
= |\Lambda|^{-1}
 \, \varepsilon_{\rm b} - \delta_{\underline{k}_{1}, 
\underline{k}_{2}} (zt)^{-1} e_1
\, \varepsilon_{\underline{k}_{2} - \underline{k}_{\rm b}} \ 
\end{equation}
\end{mathletters}
 We use the notation  $e_i:= \langle 
\Theta(\varepsilon_{\rm F}-\varepsilon_{\underline k})
\varepsilon_{\underline k}^i \rangle_{\rm BZ}$ ($\Theta(\varepsilon)$
is the Heaviside function).
The elements in (\ref{xelem})
 are obtained with the help of Wick's theorem since
$|{\cal N}'\rangle$ is a simple Slater determinant.
In (\ref{xelemb}) and (\ref{xelemc}), we distinguish the part
coming from the motion of the $\uparrow$-electrons and the part
coming from the motion of the $\downarrow$-electron.
The expression $(zt)^{-1} e_1
\, \varepsilon_{\underline{k}_{2} - \underline{k}_{\rm b}}$
in (\ref{xelemc}) is obtained from
\begin{eqnarray}\nonumber
&&-(2\pi)^{-d}\int\limits_{\underline{k}_1 \in {\rm BZ} \setminus {\rm FS}}
\varepsilon_{\underline{k}_2-\underline{k}_{\rm b}- \underline{k}_1}
d^dk_1 = 
\\ \label{isotrop}
&& \qquad \frac{1}{zt}\varepsilon_{\underline{k}_2-\underline{k}_{\rm b}}
\underbrace{
(2\pi)^{-d}\int\limits_{\underline{k}_1 \in {\rm BZ} \setminus {\rm FS}}
\varepsilon_{\underline{k}_1} d^dk_1}_{e_1:=} \ .
\end{eqnarray}
This relation
holds for all homogeneous, isotropic lattices with NN hopping only, e.g.\
square lattice, triangular lattice, kagome lattice and so on. 
The result  (\ref{isotrop}) can be found easiest by interpreting
the left hand side as convolution of 
$\varepsilon_{\underline{k}_2-\underline{k}_{\rm b}}$ and of 
$\Theta(\varepsilon_{\rm F}-\varepsilon_{\underline{k}_1})$ ,
i.e.\ as a  multiplication in real space which concerns only the NN terms.
Thus it is the multiplication with a constant 
$\langle a_{\underline{j}}^{\phantom{+}} a_{\underline{i}}^+\rangle 
= -e_1/(zt)$. The sites $\underline{i}$ and
$\underline{j}$ are arbitrary adjacent sites since all
 bonds are equal due to the
required homogeneity and spatial isotropy.

On the basis of (\ref{xelem}) the matrix inversion can be rephrased as 
\begin{equation}
\label{invert}
(\omega {\bf P} - {\bf L})^{-1} = 
\left({\bf d}^{-1} + (\omega - \varepsilon_{\rm b}) \, 
{\underline v} \, {\underline v}^{+}\right)^{-1} \ 
\end{equation}
with the constant vector ${\underline v}=|\Lambda|^{-1/2}$ and the
diagonal matrix ${\bf d}_{{\underline k}_{1} {\underline k}_{2}} = 
\delta_{{\underline k}_{1} {\underline k}_{2}}  f({\underline k}_2)$ with
\begin{equation}
\label{fdef}
f(\underline k) := [n(\omega -\varepsilon_{\underline k}) + 
e_1 (1 + (zt)^{-1} 
\varepsilon_{\underline k -\underline{k}_{\rm b}})]^{-1} \ .
\end{equation}
Note that the dyadic product ${\underline v} \, {\underline v}^{+}$
provides a $\delta|\Lambda| \times \delta|\Lambda|$ matrix
with the constant matrix element $|\Lambda|^{-1}$.

Expanding the right hand side of (\ref{invert}) in terms
of ${\underline v} \, {\underline v}^{+}$ and resummation in terms of
\begin{mathletters}
\begin{eqnarray}
h(\omega)&:=&  {\underline v}^{+} {\bf d}\, {\underline v}\nonumber \\
& =&  (2\pi)^{-d} \int\limits_{\underline k \in {\rm BZ} 
\setminus {\rm FS}}
f(\underline k) d^dk \nonumber \\
&=&  \langle \Theta(\varepsilon_{\underline k} - \varepsilon_{\rm F}) \, 
f(\underline k) \rangle_{\rm BZ}
\label{hdef0}
\end{eqnarray}
 yields
\begin{eqnarray}
&&(\omega {\bf P} - {\bf L})^{-1} \nonumber \\
 && = ({\bf d}^{-1} + (\omega - \varepsilon_{\rm b})
{\underline v} {\underline v}^{+} )^{-1} \nonumber \\
 && = {\bf d} \, ( {\bf 1} + 
(\omega - \varepsilon_{\rm b}) {\underline v} {\underline v}^{+} 
{\bf d})^{-1} \nonumber \\
 && = {\bf d} \, ( {\bf 1} - 
(\omega - \varepsilon_{\rm b}){\underline v}{\underline v}^{+} {\bf d}
 + (\omega - \varepsilon_{\rm b})^{2} {\underline v}{\underline v}^{+} 
{\bf d} {\underline v} {\underline v}^{+} {\bf d}
 + \ldots) \nonumber \\
 && = {\bf d} - (\omega - \varepsilon_{\rm b}) \, 
{\bf d} {\underline v} {\underline v}^{+} {\bf d} 
 \, (1 + (\omega - \varepsilon_{\rm b}) \, h(\omega))^{-1} \ .
\label{funda0}
\end{eqnarray}
The matrix elements thus read
\begin{eqnarray} \nonumber
&&\left((\omega{\bf P}-{\bf L})^{-1}\right)_{{\underline k}_{1}, 
{\underline k}_{2}} = 
\delta_{{\underline k}_1, {\underline k}_2} \, f({\underline k}_1) 
\\
&& \qquad 
- \frac{\omega - \varepsilon_{\rm b}}{1+ (\omega - \varepsilon_{\rm b}) 
h(\omega)} \cdot
\frac{f({\underline k}_1) \, f({\underline k}_2)}{|\Lambda|} \ .
\label{funda1}
\end{eqnarray}
\end{mathletters}
From (\ref{funda1}) we read off that $(\omega {\bf P} - {\bf L})$
is singular for 
\begin{equation}
\label{res0}
0=1 + (\omega - \varepsilon_{\rm b}) \, h(\omega)\ .
\end{equation}
The trick to reduce dyadic perturbations to simple divisions is
commonly known under the name `Householder method'
in the numerics of matrices.
This extremely simple result is derived here for all Bravais lattices,
e.g.\ the square lattice, the triangular lattice, but not
for the honeycomb lattice or the kagome lattice.
The restriction to Bravais lattices enters since we implicitly
assume that there is one eigen state for each value of ${\underline k}$
in the one-particle Hamiltonian. But it will be shown in  
appendix \ref{app:unfrustrated} that identical formulae
apply for general
unfrustrated lattices. Similar formulae can be found for frustrated
non-Bravais lattices, for instance the kagome lattice in
appendix \ref{app:kagome}.

In appendix \ref{app:tri} it is explained that 
computing $h(\omega)$ for non-bipartite lattices requires explicit
integration over the momenta.
For lattices  where the band minimum $\varepsilon_{\rm b}$ is reached at 
${\underline k}_{\rm b}={\underline 0}$ further 
simplification is possible. 
The band minimum is found at $\underline{k}_{\rm b}=\underline{0}$
in particular for bipartite lattices 
where one may choose $t>0$ without loss of generality.
Then the term $\varepsilon_{\underline k -\underline{k}_{\rm b}}$ in 
(\ref{fdef})
reduces to the unshifted dispersion and the whole integration
in (\ref{hdef0}) can be written as integration over the
density of states (DOS) $\rho(\varepsilon)$.
\begin{mathletters}
\label{kb0def1}
\begin{equation}
\label{hdef}
h(\omega) = \int\limits_{\varepsilon_{\rm F}}^{\varepsilon_{\rm t}} 
\frac{\rho (\varepsilon) \, d\varepsilon}{n \omega + e_{\cal N}
- \gamma \varepsilon} = \frac{1}{\gamma} \ G(\Omega)
\end{equation}
\begin{eqnarray}
\label{Gdef}
G(y) &:=& \int\limits_{\varepsilon_{\rm F}}^{\varepsilon_{\rm t}} 
\frac{\rho(\varepsilon) \, d\varepsilon}{y - \varepsilon} \\
 \gamma &:=& n - e_1/zt \label{gamdef}
\\
\Omega &:=& (n \omega + e_1)/\gamma
\end{eqnarray}
\end{mathletters}

We will call the ansatz deduced from the subspace
given in (\ref{varia}) RES0. It leads to the singularity condition
(\ref{res0}) or to its generalizations for non-Bravais lattices.

Once the energy $\omega$ is found
from (\ref{res0}) for a given Fermi energy $\varepsilon_{\rm F}$ 
the spin flip energy for the whole process of taking
one $\uparrow$-e$^-$ out at the Fermi level and inserting it as
$\downarrow$-e$^-$ at the lowest possible energy is given by
\begin{equation}
\label{spinflip2}
\Delta e_\infty = \omega -\varepsilon_{\rm F} \ .
\end{equation}
A critical doping $\delta_{\rm cr}$ is found where this spin flip
energy vanishes.

\subsection{Case $U<\infty$: ansatzes RES1, RES2, and RES3}
Besides the calculation of variational upper bounds for spin flip
energies and resulting critical dopings it is our aim to determine
critical interaction values $U$. 

For $U<\infty$ we have to include states with double occupancy.  The
easiest way to do so is to include a local double occupancy
\cite{gutzw65,shast90a}. This is done in the ansatz RES1 by adding to
the states defined in (\ref{varia}) the state 
\begin{equation}
\label{varib}
| \Psi_1 \rangle
 :=  |\Lambda|^{-1/2}\sum\limits_{\underline{i}}
 \exp(i\underline{k}_{\rm b} \underline{i})
a_{\underline{i}\uparrow}^{+}
a_{\underline{i}\uparrow}^{\phantom{+}}
a_{\underline{i}\downarrow}^{+} | {\cal N}' \rangle \ .
\end{equation}
This ansatz contains the nearest neighbor ansatz NN (\ref{ansatz:gw})
(and the Gutzwiller ansatz) for $U<\infty$.
Again we want to compute the resolvent (\ref{resolv}).
To do so the parts computed in the previous
subsection can be used again. The matrices for RES1
have the block structure
\begin{mathletters}
\label{block1}
\begin{equation}
{\bf P} = \left(
\begin{array}{c|c}
  {\bf P_1} & {\underline 0}^+        \\
  \hline
  {\underline 0}  & P_2
\end{array}  
\right) \ \
\omega{\bf P} - {\bf L} = \left(
\begin{array}{c|c}
  {\bf D_1} & {\underline N}        \\
  \hline
  {\underline N}^+   & D_2
\end{array}  
\right)
\end{equation}
\begin{equation}
(\omega{\bf P} - {\bf L})^{-1} = \left(
\begin{array}{c|c}
  {\bf B_1} & {\underline M}        \\
  \hline
  {\underline M}^+   & B_2
\end{array}  
\right)\ .
\end{equation}
\end{mathletters}
The matrices ${\bf P_1}$ and ${\bf D_1}$ are the same as in (\ref{funda1})
at $U=\infty$. The null vector ${\underline 0}$ in ${\bf P}$ comes from 
the fact
that the state with double occupancy
$| \Psi_1 \rangle$ is orthogonal to the states without double
occupancy $| \Phi_{\underline k} \rangle$. The other matrix elements are
again found by Wick's theorem
\begin{mathletters}
\label{xelem2}
\begin{eqnarray}
P_2 &=& n
\\
D_2 &=& n(\omega - U) + e_1 - \varepsilon_{\rm b} \left(n^{2} - 
\left( e_1/(zt)
\right)^{2} \right)
\\
N_{\underline k} &=&   -|\Lambda|^{-1/2} \left( 
n (\varepsilon_{\rm b} - \varepsilon_{\underline k}) + 
e_1 (1 +\varepsilon_{\underline k - \underline{k}_{\rm b}}/(zt)) \right) 
\end{eqnarray}
\end{mathletters}
Since we are at present only interested in the singularity condition
it is sufficient to compute one of the elements of 
$\omega{\bf P}- {\bf L}$. The easiest is $B_2$, for which an argument
similar to the one leading to (\ref{funda0}), yields
\begin{equation}
B_2 = (D_2 - {\underline N}^+ {\bf D}_{{\bf 1}}^{-1} 
{\underline N})^{-1}\ . 
\end{equation}
Thus the singularity condition simply reads
\begin{equation}
\label{singcond}
0 \doteq D_2 - {\underline N}^+ {\bf D}_{{\bf 1}}^{-1} {\underline N}
\end{equation}
Now it is advantageous that ${\bf D_1}^{-1}$ is already given in
(\ref{funda1}). Inserting (\ref{xelem2}) one obtains after some
cancellations
\begin{equation}
\label{res1}
\omega - \varepsilon_{\rm b} - n U (1 + 
(\omega - \varepsilon_{\rm b})  h(\omega)) \doteq 0 \ .
\end{equation}
Equation (\ref{res1}) is as simple as (\ref{res0}) and enables
us to calculate critical $U$ values explicitly. 
Setting $\omega=\varepsilon_{\rm F}$ in (\ref{res1}), which according to
(\ref{spinflip2}) corresponds to vanishing spin flip energy, 
renders $U_{\rm cr}$ directly accessible:
\begin{equation}
\label{ures1}
U_{\rm cr}^{\rm RES1} (\delta) = \frac{\varepsilon_{\rm F} - 
\varepsilon_{\rm b}}{(1- \delta) \, 
[1+ (\varepsilon_{\rm F} - \varepsilon_{\rm b}) \, h(\varepsilon_{\rm F})]} \ .
\end{equation}

It turns out, however, that the values for $U_{\rm cr}$ from
(\ref{ures1}) are not very good close to half-filling $n=1$ where
antiferromagnetic exchange processes are important.  These are not
accounted for in (\ref{varib}). They are considered, at least
to a certain extent, in the ansatz RES2 by using
\begin{equation}
\label{varic}
| \Psi_2 \rangle
 :=  |\Lambda|^{-1/2}\sum\limits_{<\underline{i}\underline{j}>}
 \exp(i\underline{k}_{\rm b} \underline{i})
a_{\underline{i}\uparrow}^{+}
a_{\underline{j}\uparrow}^{\phantom{+}}
a_{\underline{i}\downarrow}^{+} | {\cal N}' \rangle 
\end{equation}
instead of $| \Psi_1 \rangle$ as extension of the RES0 subspace.

The block structure (\ref{block1}) remains the same and so does
the singularity condition (\ref{singcond}). Only the matrix elements
are modified
\begin{mathletters}
\label{xelem3}
\begin{eqnarray}
&&P_2 = (e_1^2 + \delta e_2)/t^2
\\
\nonumber
&&D_2= ((e_1^2 + \delta e_2)(\omega - U) + 
e_1 e_2 +\delta e_3  
\\ \label{d2res2}
&& \qquad - \varepsilon_{\rm b} 
\left(e_1^{2} -  e_1e_3/(zt)^2\right))/t^2
\\
&&N_{\underline k} = \nonumber \\
&&   |\Lambda|^{-1/2} \left( e_1 
(\varepsilon_{\rm b} - \varepsilon_{\underline k}) + 
e_2\left(1 +  \varepsilon_{\underline k - 
\underline{k}_{\rm b}}/(zt)\right) \right)/t \ .
\end{eqnarray}
\end{mathletters}
The explicit expression resulting now
 from (\ref{singcond}) is less transparent 
than (\ref{res1}) since no cancellations occur. We focus here on the most
important case ${\underline k}_{\rm b}={\underline 0}$. 
In addition to the definitions 
(\ref{kb0def1}) we use
\begin{mathletters}
\begin{eqnarray}
\gamma' &:=& e_1 - e_2/(zt)
\\
\Omega_{\rm b} &:=& (e_1\varepsilon_{\rm b} + e_2)/\gamma'
\\
\label{ydef}
y &:=& (\gamma'/\gamma)\left(\delta + 
(\Omega_{\rm b}-\Omega)G(\Omega)\right)
\end{eqnarray}
\end{mathletters}
and obtain from (\ref{singcond})
\begin{eqnarray}
D_2 &\doteq& \sum\limits_{{\underline k}_1,{\underline k}_2 \in 
{\rm BZ}\setminus{\rm FS}}
N_{{\underline k}_1}^+ ({\bf D_1}^{-1})_{{\underline k}_1,
{\underline k}_2}N_{{\underline k}_2}
\nonumber \\ \nonumber 
&=& \gamma'[(\Omega_{\rm b} - \Omega) y + \frac{\gamma'}{\gamma}
(\delta\Omega_{\rm b} + e_1)] 
\\
&& \quad - y^2 
\frac{\omega - \varepsilon_{\rm b}}{1+(\omega - 
\varepsilon_{\rm b})h(\omega)}
\label{res2}
\end{eqnarray}
from which $U_{\rm cr}$ can easily be determined. The value
$U_{\rm cr}$ appears only in
$D_2$, see (\ref{d2res2}). The results of RES2 (\ref{res2}) generically
lead to $U_{\rm cr}\propto 1/\delta$ on vanishing doping. In this sense it
represents an important improvement over RES1 (\ref{res1}). For explicit
results we refer the reader to the next section.

At last in RES3, we generalize the variational states with double occupancy
like (\ref{varib}) and (\ref{varic}) in the same manner as we 
generalized the states without double occupancy in (\ref{varia})
\begin{mathletters}
\label{varid}
\begin{equation}
| \Psi_{\underline{k}} \rangle
 := {\cal B}_{\underline{k}} | {\cal N}' \rangle 
\end{equation}
\begin{equation}
{\cal B}_{\underline{k}}  := 
|\Lambda|^{-1/2}\sum\limits_{\underline{i}}
 \exp(i(\underline{k}_{\rm b} + \underline{k})\underline{i})
a_{\underline{i}\uparrow}^{+}
c_{\underline{k}\uparrow}^{\phantom{+}} 
a_{\underline{i}\downarrow}^{+}  \ ,
\end{equation}
\end{mathletters}
where now the admissible values of $\underline{k}$ are all vectors
inside the Fermi sphere (FS).
Note that the extension RES3 contains both RES1 and RES2.
The block structure of the resulting problem is 
similar to the one in (\ref{block1}). The difference is that
all blocks are now macroscopically large
\begin{mathletters}
\label{block2}
\begin{equation}
{\bf P} = \left(
\begin{array}{c|c}
  {\bf P_1} & {\bf 0}^+        \\
  \hline
  {\bf 0}   & {\bf P_2}
\end{array}  
\right), \quad
\omega{\bf P} - {\bf L} = \left(
\begin{array}{c|c}
  {\bf D_1} & {\bf N}        \\
  \hline
  {\bf N}^+   & {\bf D_2}
\end{array}  
\right)
\end{equation}
\begin{equation}
(\omega{\bf P} - {\bf L})^{-1} = \left(
\begin{array}{c|c}
  {\bf B_1} & {\bf M}        \\
  \hline
  {\bf M}^+   & {\bf B_2}
\end{array}  
\right)\ .
\end{equation}
\end{mathletters}
The matrix elements and details of the evaluation are
given in the appendix \ref{app:res3}. The main problem
is that one has to find a tractable condition for 
\begin{equation}
\label{B2def}
{\bf B_2}^{-1} = {\bf D_2}- {\bf N}^+{\bf D_1}^{-1}{\bf N}
\end{equation}
to be singular. But with expansion tricks similar to the ones
used above this obstacle can be overcome.
For bipartite lattices a relatively simple final formula
is found (\ref{bound4}). An evaluation for the
triangular lattice (appendix \ref{app:tri}) and the
kagome lattice (appendix \ref{app:kagome})  is possible as well.

\section{Results for various lattices} \label{sect:resvarlat}
\subsection{Square lattice} \label{subs:square}

The square lattice represents the simplest bipartite lattice structure
in two space dimensions and has therefore been at the center of interest
in most of the publications dealing with the variational investigation of 
Nagaoka 
stability\cite{shast90a,basil90,gebha91a,linde91,hanis93,wurth95,wurth96}. 
The energy band is given by
\begin{equation} \label{quad_dispers}
\varepsilon_{\Box} (\underline{k}) = -2t \, (\cos k_{x} + \cos k_{y}) \ ,
\end{equation}
with the lattice spacing set to 1. The DOS 
$\rho_{\Box} (\varepsilon)$ which is depicted in fig.\ \ref{fig:quad_1}(a)
can be expressed by a complete elliptic integral of the first kind (see
appendix \ref{app:dos}). For positive hopping matrix element $t$ the 
lower band edge is reached at $\underline{k}_{\rm b} =
\underline{0}$ while the maxima of the band structure are
located at the corners of the square shaped first Brillouin zone
($\underline{k}_{\rm t}= (\pm \pi, \pm \pi)$). The logarithmic
van Hove singularity at $\varepsilon = 0$ corresponds to the saddle points of
the dispersion (\ref{quad_dispers}). The symmetric shape of the DOS 
with respect to $\varepsilon = 0$ reflects the 
particle-hole symmetry of the Hubbard model on the square lattice. In the
following we make use of this symmetry and consider only the case of
a less than half filled lattice ($0 \leq n \leq 1$) and $t>0$.
\begin{figure}[htb]
  \setlength{\unitlength}{1cm}
 \begin{picture}(8,13.8)
   \includegraphics{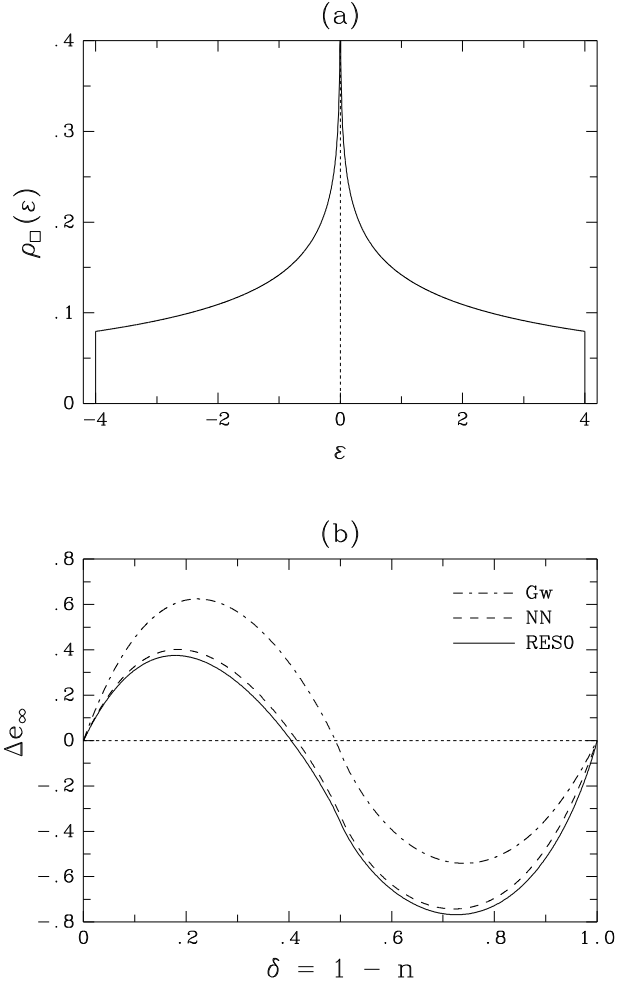}
 \end{picture} \par
 \caption{ \label{fig:quad_1} 
   (a) DOS for the square lattice ($t=1$), (b) spin flip energy at
   $U=\infty$ as a function of the hole density for Gw, NN and RES0 
   on the square lattice ($t=1$).}
\end{figure} 

Fig.\ \ref{fig:quad_1}(b) shows the spin flip energies at 
$U=\infty$ resulting from the variational criteria discussed in the previous 
sections  as a function of $\delta$. 
The  Gutzwiller wave function ((\ref{ansatz:gw}) with $f=0$)
gives a critical hole density $\delta_{\rm cr} = 0.4905$ for the instability 
of the Nagaoka state\cite{shast90a}. For the variational 
ansatz (\ref{ansatz:gw}) including nearest neighbor hopping processes 
of the majority spins (finite $f$), the spin flip energy is considerably
lowered and the critical hole density decreases to $\delta_{\rm cr} = 0.4155$.
The evaluation the variational state RES0, 
which contains \textit{all} spin-up 
hopping terms of the Basile-Elser type, leads to $\delta_{\rm cr} = 0.4045$. 
Thereby we reproduce up to the fifth digit our result obtained in 
\cite{hanis93} where we took into account 
hopping processes over a distance of up to four lattice spacings. 

The fact that the reduction of the spin flip energy in 
fig.\ \ref{fig:quad_1}(b) is mainly due to the nearest neighbor 
term demonstrates the overwhelming
importance of \textit{local} polarizations of the spin up Fermi sea for
the instability of the Nagaoka state.
 The resolvent method  treats implicitly an \textit{infinite} 
number of variational parameters and
makes it  
\begin{figure}[htb]
  \setlength{\unitlength}{1cm}
 \begin{picture}(8,13.8)
   \includegraphics{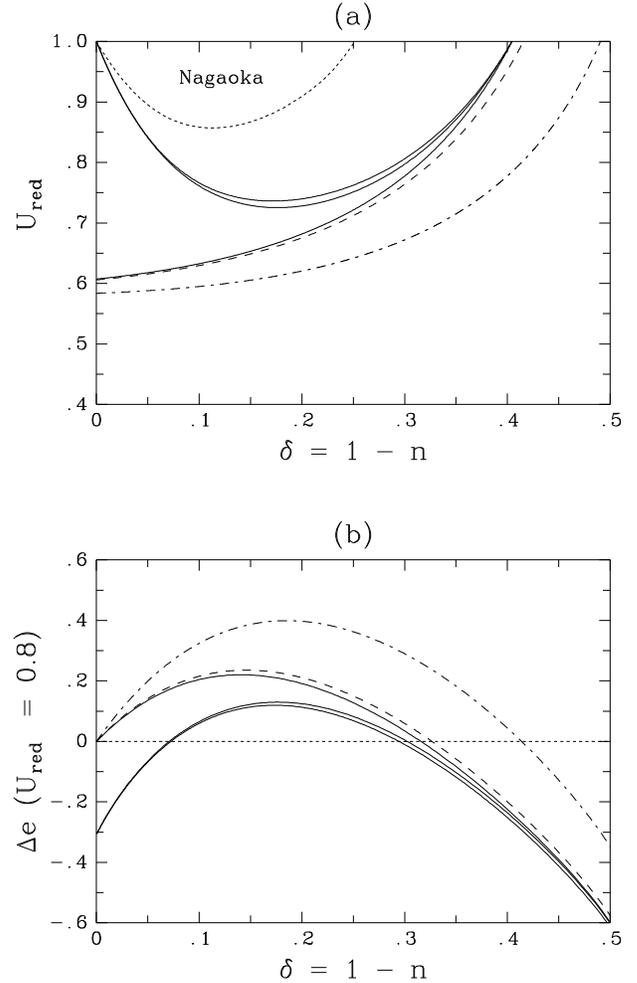}
 \end{picture} \par
 \caption{ \label{fig:quad_2} 
   (a) Phase diagram ($n<1$): Nagaoka instability lines on the square
   lattice for Gw (dashed-dotted), NN (long dashed), RES1, RES2, RES3
   (full lines, from bottom to top), and the 1100 parameter ansatz of
   Wurth et al.\cite{wurth96} (short dashed), (b) spin flip energy for
   $U_{\rm red} = 0.8$ and $t=1$ as a function of the hole density for Gw
   (dashed-dotted), NN (dashed), RES1, RES2, and RES3 (full lines,
   from top to bottom).}
\end{figure} \noindent
 possible to investigate the full Basile-Elser ansatz for the first time
in the thermodynamic limit. Compared to the iterative 
method used in \cite{hanis93} it has the remarkable advantage that the lowest 
possible spin flip energy in a given subspace can be calculated 
\textit{without} explicit 
knowledge of the corresponding state. 
As we will see in section E, 
it is not generally true that the best value for $\delta_{\rm cr}$ 
within the Basile-Elser subspace can be obtained by restricting the spin 
up hopping processes to a small cluster centered at the position 
of the flipped spin.

Fig.\ \ref{fig:quad_2}(a) shows the Nagaoka instability lines in the 
phase diagram for the Gutzwiller single spin flip (Gw), the 
nearest neighbor ansatz (NN) (\ref{ansatz:gw}) 
 as well as for the wave 
functions RES1, RES2, and RES3 evaluated by means of the 
resolvent method. The $\uparrow$-hopping terms appear to be much 
less efficient in suppressing the Nagaoka state if the hole density is 
small (because most of the sites near the flipped spin are already occupied by
a spin-up electron) and  the on-site repulsion $U$ is finite (because the 
terms  all exclude double occupancies at the down spin position). Since the
Gutzwiller projector (with $g > 0$)  represents the only term 
contained in RES1 which is relevant for $U<\infty$, the critical on-site 
repulsion near half filling is only slightly increased and $U_{\rm cr}$ remains 
finite for $\delta= 0$. A remarkable improvement is obtained by allowing
for nearest neighbor exchange processes and thereby taking 
into account the antiferromagnetic 
tendency  of the nearly half-filled Hubbard model. This is embodied in the 
ansatz RES2. For a constant non-zero value of the DOS 
at the upper band edge it leads to the asymptotic
behavior $U_{\rm cr, red} (\delta) = 1 - {\cal O}(\delta)$ for $\delta
\rightarrow 0$. This implies the instability of the Nagaoka state for 
all finite values of $U$ in this limit. Fig.\ \ref{fig:quad_2}(b) shows 
that the optimum spin flip energy for RES2 plotted as a function 
of the hole density for a fixed finite value of $U$ approaches a finite 
negative value of the order $t^2/U$ at half filling while it vanishes for 
all wave functions containing only the Gutzwiller projector. 

The asymptotic behavior for $\delta \rightarrow 0$ of the spin flip energy 
and of the Nagaoka instability line $U_{\rm cr} (\delta)$ is not affected by the
extension of the Hilbert subspace to the full resolvent ansatz RES3. As
for $U=\infty$ the \textit{local} terms play the most important role in
destabilizing Nagaoka ferromagnetism.
With increasing hole density exchange processes become less important
and the Nagaoka instability lines for RES2 and RES3 approach the one obtained 
for RES1. Since \textit{all} RES wave functions differ only in the
subspace with double occupancies the 
corresponding instability lines end up with a diverging on-site 
repulsion $U_{\rm cr}$ at  the critical hole density 
$\delta_{\rm cr} = 0.4045$  obtained for RES0.

Fig.\ \ref{fig:quad_2}(a) displays also the best known variational 
bound for the Nagaoka stability regime on the square lattice computed by 
Wurth et al.\ \cite{wurth96}. The corresponding state contains 1100 terms, 
most of them describing excitations
of the spin-up Fermi sea with up to two particle-hole pairs located within
a 9$\times$9 plaquette around the down spin position. The critical hole
density obtained with this variational wave function is $\delta_{\rm cr} = 0.2514$
and the minimum critical on-site repulsion is $U_{\rm cr}^{\rm min}/t = 77.74$
(RES3: $U_{\rm cr}^{\rm min}/t = 36.21$). Comparing these results one should keep
in mind that the resolvent method allows to derive analytic expressions 
for the Nagaoka instability line $U_{\rm cr} (\delta)$, at least for RES1 and 
RES2, while the calculation of the phase boundary for the 1100 parameter state 
requires an immense numerical effort.

\subsection{Square lattice with next-nearest neighbor hopping
\label{subs:tprime}}

Extending the Hamiltonian (\ref{def:model}) by taking next-nearest 
neighbor hopping processes of the electrons into account and introducing
a corresponding hopping amplitude $t'$  allows to \textit{create} a 
particle-hole asymmetry of the DOS. Variation of the ratio $t'/t$ 
makes it possible to simulate a continuous ``transition'' between a bipartite 
and a non-bipartite lattice. In this subsection we investigate how this 
transition affects the stability of the Nagaoka state with respect to a 
Gutzwiller single spin flip on the square lattice. Furthermore we will give
a perturbation argument  for $|t| \ll |t'|$.

The band dispersion of the so-called $t$-$t'$-$U$ model on the square lattice
is given by
\begin{equation} \label{ttstrich_dispers}
\varepsilon_{t-t'} (\underline{k}) = -2t \, (\cos k_x + \cos k_y)
-4t' \cos k_x \cos k_y \ .
\end{equation}
For $t,t'>0$ the lower band edge $\varepsilon_{\rm b} 
= -4(t+t')$ is reached at $\underline{k}_{\rm b} 
= \underline{0}$. The maxima of the band structure are located at the 
corners  of the
Brillouin square for $t' < t/2$ and at the edge centers for $t' > t/2$, 
respectively. Exactly for $t'=t/2$ the maximum single particle 
energy $\varepsilon_{\rm t} = 2t$ is reached at the whole border 
of the Brillouin zone. This leads to a nesting situation 
and to the largest possible partice-hole aymmetry with a diverging DOS
at the upper band edge. For $t' > t/2$ local minima of the band structure 
develop at the corners of the Brillouin zone leading to a step in the DOS.
In the limit $t/t' \rightarrow 0$ the
single particle energy at these $\underline{k}$-points reaches 
the lower band edge. The calculation of the DOS 
$\rho_{t-t'} (\varepsilon)$ requires in general a numerical 
$\underline{k}$-integration. Only for $t'=t/2$ it is possible to map 
$\rho_{t-t'} (\varepsilon)$ on the DOS for $t=0$ and hence on a complete
elliptic integral (see appendix \ref{app:dos}):
\begin{equation} \label{zust_ttstrich}
\rho_{t-t'} (\varepsilon) = \left(1 - \frac{\varepsilon}{2t}\right)^{-1/2}
\ \ \rho_{\Box} \left(2t \sqrt{1 - \frac{\varepsilon}{2t}} \right) 
\ .
\end{equation}

The symmetry of the Nagaoka stability regime with respect to half filling 
found in the ``pure'' Hubbard model is destroyed if the next-nearest neighbor 
hopping $t'$ is switched on. In analogy to the non-bipartite triangular 
and kagome lattices (see \cite{hanis95} and sect.\ \ref{subs:tri} in this
paper) one should expect that the tendency towards saturated ferromagnetism 
increases for more than half filling and decreases for $n <1$. 
The RES ansatzes with the reduction to DOS integrals cannot be used for the
$t$-$t'$ model since the $t'$-hops go beyond nearest neighbor hopping.

The calculation of the optimum spin flip energy for the Gutzwiller ansatz
((\ref{ansatz:gw}) with $f=0$)
 requires additional effort for the $t$-$t'$-$U$ model 
due to the more complicated structure of the  band 
dispersion (\ref{ttstrich_dispers}). The kinetic energy of the 
flipped spin no longer depends only on $\varepsilon_{\rm b}$ but
also on the corresponding \textit{momentum} $\underline{k}_{\rm b}$.
For $t,t' > 0$ (i.e.\ for less than half filling) we find 
$\underline{k}_{\rm b} = \underline{0}$ as for $t'=0$, whereas
for  $t,t' < 0$ (i.e.\ for more than half filling) we choose 
$\underline{k}_{\rm b} = (\pi, \pi)$ for $t'/t \leq 1/2$ and 
$\underline{k}_{\rm b} = (\pi, 0)$ for $t'/t > 1/2$. 

\begin{figure}[htb]
 \setlength{\unitlength}{1cm}
 \begin{picture}(8,13.8)
   \includegraphics{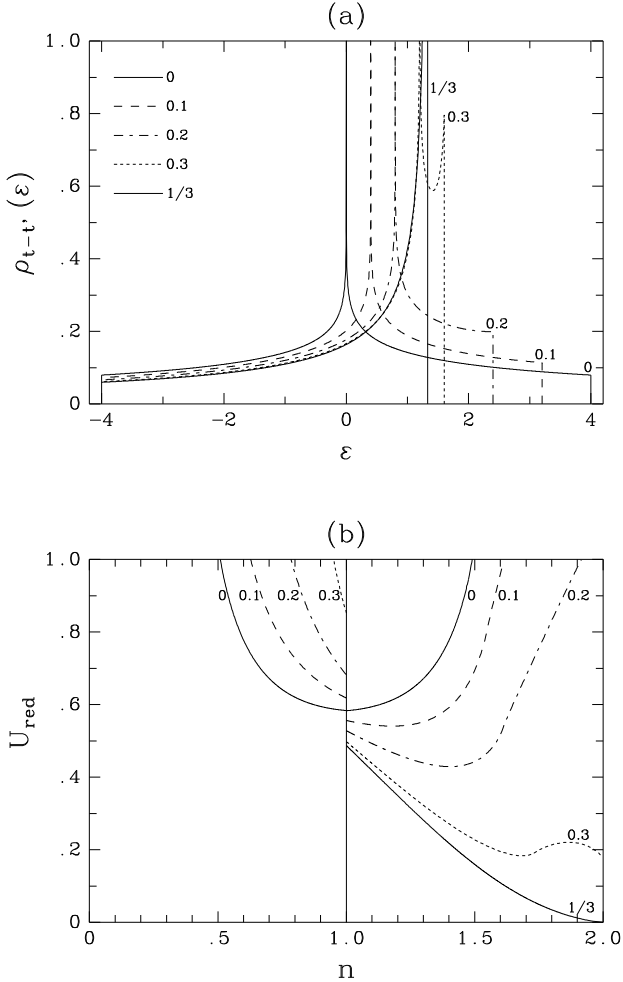}
 \end{picture} \par
\caption{ \label{fig:tstrich_1} $t$-$t'$-$U$ model on the square lattice for 
$|t'| \leq |t|/2:$ $|t'| = 1 - |t| = 0,\, 0.1,\, 0.2,\, 0.3,\,
 1/3$:
(a) DOS $\rho_{t-t'}(\varepsilon)$ for $t, t' > 0$ ($\rho_{t-t'}(\varepsilon)$
    for $t, t' < 0$ is obtained by $\varepsilon \leftrightarrow -\varepsilon$),
(b) Nagaoka instability lines for a Gutzwiller single spin flip (the curves 
    for $n<1$ correspond to $t, t' >0$ whereas the curves for $n>1$ 
    correspond to $t, t'<0$).}
\end{figure}

\begin{figure}[htb]
 \setlength{\unitlength}{1cm}
 \begin{picture}(8,6)
   \includegraphics{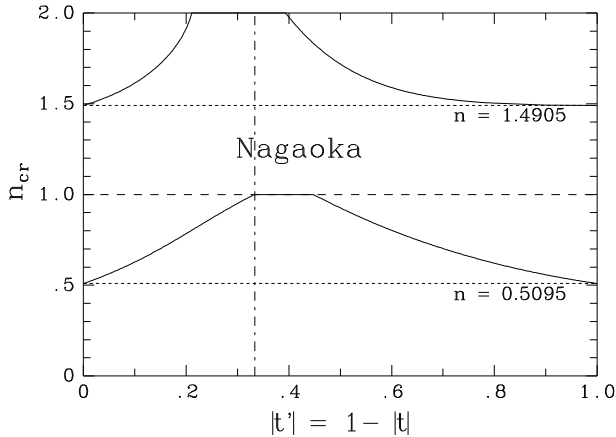}
 \end{picture} \par
 \caption{ \label{fig:tstrich_3} Critical densities for the Nagaoka instability
at $U=\infty$ for a Gutzwiller single spin flip on the square lattice as a
function of $|t'| = 1 - |t|$. Between the two full lines the Nagaoka state
is found to be possibly stable. The dashed-dotted line marks the singular case
$|t'/t|=1/2$ where the particle-hole asymmetry reaches its maximum.}
\end{figure}

Fig.\ \ref{fig:tstrich_1} shows the DOS for the
 \noindent
$t$-$t'$-$U$ model on the square lattice and the corresponding Nagaoka
instability lines in the phase diagram for various ratios $t'/t \leq 1/2$.
We set $|t| + |t'| = 1$ so that the lower band edge is always at
$\varepsilon_{\rm b} = -4$. Increasing $t'/t$ leads to a
lower DOS at $\varepsilon_{\rm b}$ and a higher DOS 
at $\varepsilon_{\rm t}$, while  the logarithmic singularity at 
$\varepsilon = 4 t'$ approaches the upper band edge. The maximum particle
hole asymmetry is reached at $|t'/t| = 1/2$ (i.e.\ $|t'|=1/3$) where the DOS
(\ref{zust_ttstrich}) diverges like 
$(\sqrt{\varepsilon_{\rm t} - \varepsilon} 
\, \cdot \, |\log (\varepsilon_{\rm t} - \varepsilon)|)^{-1}$
for $\varepsilon \approx \varepsilon_{\rm t}$.
The Nagaoka stability region for less than half filling shrinks as
$t'/t$ is increased and disappears at $t'/t = 1/2$ (Figs.\ 
\ref{fig:tstrich_1}(b), \ref{fig:tstrich_3}). On the other hand, it expands
rapidly for $n>1$, especially in the limit $n=2$. At $t'/t =1/2$  the 
Nagaoka state is stable 
towards a Gutzwiller single spin flip for all $U>0$ in this limit. Even
the slope of the Nagaoka instability line $U_{\rm cr} (n)$ vanishes at $n=2$. 

\begin{figure}[htb]
 \setlength{\unitlength}{1cm}
 \begin{picture}(8,13.8)
   \includegraphics{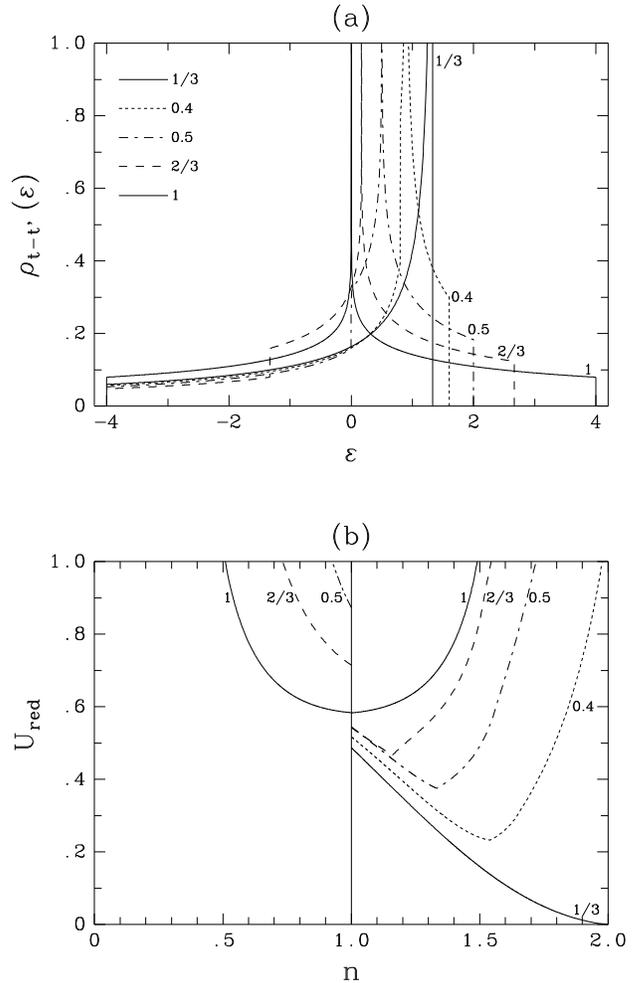}
 \end{picture} \par
\caption{ \label{fig:tstrich_2} $t$-$t'$-$U$ model on the square lattice for 
$|t'| \geq |t|/2:$ $|t'| = 1 - |t| = 1/3,\, 0.4,\, 0.5,\, 2/3,\, 1$:
(a) DOS $\rho_{t-t'}(\varepsilon)$ for $t, t' > 0$ ($\rho_{t-t'}(\varepsilon)$
    for $t, t' < 0$ is obtained by $\varepsilon \leftrightarrow 
    -\varepsilon$), 
(b) Nagaoka instability lines for a Gutzwiller single spin flip (the curves 
    for $n<1$ correspond to $t, t' >0$ whereas the curves for $n>1$ 
    correspond to $t, t'<0$).}
\end{figure}

If one increases the ratio $t'/t$ beyond 1/2, the logarithmic 
singularity in the DOS is gradually shifted back towards 
$\varepsilon = 0$ and the shape of $\rho_{t-t'} (\varepsilon)$ becomes more 
and more symmetric (fig.\ \ref{fig:tstrich_2}(a)).
 \noindent
Nevertheless  the step at $\varepsilon = 4t'(1-t/t')$ remains present for all
$t/t' >0$. The DOS at $t=0$ is identical to 
$\rho_{\Box} (\varepsilon)$, which reminds us that the $t'$-$U$ model with
suppressed nearest neighbor hopping consists of \textit{two} 
completely decoupled square lattices.

At $t'=t$ the Nagaoka stability region in the phase diagram is found to
be still very asymmetric with respect to $n=1$ (fig.\ \ref{fig:tstrich_2}(b)).
A further increase of $t'/t$ makes the phase boundaries above and below
half filling approach the ones obtained at $t=0$.
Within our variational calculations, the local stability of the 
saturated ferromagnetic state is identical in both limiting cases
$t' = 0$ and $t =0$, but see the
perturbative argument below. The step in the DOS, however, leads 
to a cusp in the Nagaoka stability line $U_{\rm cr} (n)$ for all $|t/t'| <2$.
For $t \rightarrow 0$, this cusp approaches $n=1$ and 
$dU_{\rm cr}/dn|_{n=1+}$ is discontinuous at $t=0$.
This represents a qualitative
difference to the limit $t' \rightarrow 0$. 

In fig.\ \ref{fig:tstrich_3} the upper and lower critical densities  for the
 \noindent
Nagaoka instability at $U=\infty$ are plotted as functions of
$|t'| = 1 - |t|$. Thereby we once again demonstrate the shift of the 
Nagaoka stability region towards more than half filling with increasing 
particle-hole asymmetry in the DOS. The regimes
of complete Nagaoka stability for $n>1$ ($-0.21 \leq t' \leq -0.39$) and
of complete Nagaoka instability for $n<1$ ($1/3 \leq t' \leq 0.45$) are
\textit{not} symmetric with respect to $|t'| = 1/3$. There are two different
reasons for this asymmetry. First, since the increase of the DOS 
at the lower band edge is more pronounced for $t' \searrow -1/3$ (that is, 
on the left hand side of the dashed-dotted line in fig.\ \ref{fig:tstrich_3})
than for $t' \nearrow -1/3$, also the tendency towards saturated 
ferromagnetism in the low density limit (corresponding to $n \rightarrow 2$ in 
fig.\ \ref{fig:tstrich_3}) is stronger in the former case. Second, the 
Nagaoka instability condition near half filling is essentially determined by 
the ratio $\varepsilon_{\rm t}/(zt)$, i.e.\ by the asymmetry of 
the band edges with respect to $\varepsilon = 0$. The
fact that the latter asymmetry is more pronounced for $|t'|>1/3$ than for
$|t'| < 1/3$ is responsible for the instability of the Nagaoka state 
for less than half filling on the right hand side of the dashed-dotted 
line in fig.\ \ref{fig:tstrich_3}.

In the limit $t\to 0$, a perturbative arguments gives further insight
in the stability of saturated ferromagnetism. Starting point is the observation
that at $t=0$ the square lattice decomposes into two independent
 square lattices tilted by 45$^\circ$ with hopping element $t'$.
Without any $t$ the two independent Nagaoka states on 
each sub-lattice can be oriented arbitrarily without influencing
the energy. Thus we deal with a degenerate situation and investigate by
$E^{(2)}$  (second order perturbation coefficient in $t$)
whether the parallel or the antiparallel orientation is
favored. The linear order $E^{(1)}$ vanishes for particle-hole symmetry
reasons and does not lift the degeneracy.

\begin{figure}[htb]
 \setlength{\unitlength}{1cm}
 \begin{picture}(8,6)
   \includegraphics{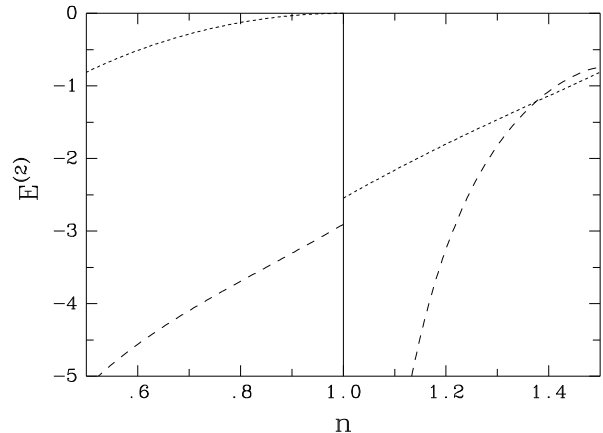}
 \end{picture} \par
 \caption{ \label{fig:tstrich_4} Second order perturbation coefficient
 $E^{(2)}$ in $t$
in units of $4t'N$. Dotted line: for parallel Nagaoka states 
(or global ferromagnetic state, see text);
 Dashed line: upper bound to $E^{(2)}$ for
antiparallel Nagaoka states (or global antiferromagnetic state).}
\end{figure}

For the parallel configuration it is straightforward to calculate $E^{(2)}$.
 Without loss of generality  we choose $t'=1/4$ and consider
$\varepsilon({\underline k}) = \varepsilon_0({\underline k}) -
 2t(\cos(k_x)\pm\cos(k_y))$
with $\varepsilon_0({\underline k}) = -\cos(k_x)\cos(k_y)$
as dispersion. The plus sign refers to $n<1,t'>0$ and the minus sign
to $n>1,t'>0$. This can be seen by means of a particle-hole transformation
and a sign transformation $c_{\underline i} \to -c_{\underline i}$ on all sites
with {\em even} $x$-coordinate.
One obtains at constant filling $E^{(2)} = - |\Lambda|/(2t') 
A_\pm(\varepsilon_{\rm F})$ with
\begin{eqnarray}\nonumber
&&A_\pm(\varepsilon_{\rm F}) =\\ \nonumber
&& \int\limits_{-\pi}^\pi \frac{d^2k}{(2\pi)^2}
(\cos(k_x)\pm\cos(k_y))^2 \delta(\varepsilon_{\rm F} + \cos(k_x)\cos(k_y))
\\ \label{hamueha}
& =& -\frac{4}{\pi^2}\left(\pm \varepsilon_{\rm F} 
K(1-\varepsilon_{\rm F}^2) - E(1-\varepsilon_{\rm F}^2)
\right)
\end{eqnarray}
yielding the dotted curves in fig.\ \ref{fig:tstrich_4}. The relation
 \noindent
(\ref{hamueha}) is found with the help of the quantities $I_n$ in
appendix A of Hanisch/M\"uller-Hartmann\cite{hanis93}; $K$ and $E$
are complete elliptic integrals. Note that the
coefficient $E^{(2)}$ is not continuous across $n=1$.

Next we assess the energy  of two antiparallel Nagaoka states on each of the
sub-lattices. Let us use $a^+_{{\underline k},\sigma}$ 
for the fermions on the
A sub-lattice and  $b^+_{{\underline k},\sigma}$ for the fermions on the
B sub-lattice. The perturbation reads then 
\begin{equation}
H_1 = -2t\sum\limits_{{\underline k}\in  {\rm MBZ}}(\cos(k_x)\pm\cos(k_y))
(a^+_{{\underline k},\sigma} b^{\phantom +}_{{\underline k},\sigma}
 + b^+_{{\underline k},\sigma} a^{\phantom +}_{{\underline k},\sigma})
\end{equation}
where MBZ is the magnetic Brillouin zone.
The second order energy lowering is
\begin{equation}
\label{perturb1}
E^{(2)} t^2 |\Lambda| = -\left\langle A\uparrow, B\downarrow \left|
H_1 (H_0-E_0)^{-1} H_1 \right| A\uparrow, B\downarrow\right\rangle ) \ .
\end{equation}
The acronyms $A\uparrow$ and $B\downarrow$ stand for the 
respective Fermi seas.
There are two processes which contribute equally to (\ref{perturb1}).
Either a fermion is shifted from $A$ to $B$ and back or a fermion is 
shifted from $B$ to $A$ and back. The latter yields explicitly
\begin{eqnarray}
&&E^{(2)} = - \frac{8}{|\Lambda|} \nonumber
\sum\limits_{{\underline k}\in  {\rm MBZ}}(\cos(k_x)\pm\cos(k_y))^2
\Theta(\varepsilon_{\rm F}-\varepsilon_0({\underline k})
\\
&& \qquad \left\langle A\uparrow \left|
a^{\phantom +}_{{\underline k},\uparrow}
(H_{0,A}-E_{0,A} - \varepsilon({\underline k}))^{-1}
a^+_{{\underline k},\uparrow}
\right|A\uparrow \right\rangle 
\\
&& = 4\int\limits \frac{d^2k}{(2\pi)^2}
(\cos(k_x)\pm\cos(k_y))^2 g_{\underline k}(\varepsilon_0({\underline k}) 
\Theta(\varepsilon_{\rm F}-\varepsilon_0({\underline k})\nonumber
\end{eqnarray}
where $g_{\underline k}$ is the one-particle Green function. Now we specify
that we work at $U=\infty$ and we {\em assume} that  the Nagaoka state 
is stable for $t=0$ at the filling considered. If the Nagaoka state is not
stable we do not need to make the present comparison anyway.
Based on our assumption, the Green function is
purely real and negative. It obeys the inequality
\begin{mathletters}
\begin{eqnarray}\label{abschaetz}
g_{\underline k}(\varepsilon_0({\underline k})) &<& 
(\varepsilon_0({\underline k})- e_{\underline k})^{-1} <0
\\ \nonumber
e_{\underline k} &:=& \langle a^{\phantom +}_{{\underline k},\uparrow}|
H_{0,A}-E_{0,A}
|a^+_{{\underline k},\uparrow} \rangle
\\ \label{approx_ska}
&=& -e_1/\delta + \varepsilon_0({\underline k}) \delta(1-(e_1/\delta)^2) \ .
\end{eqnarray}
\end{mathletters}
The  estimate (\ref{abschaetz}) corresponds to a simple Gutzwiller ansatz
\cite{shast90a} and yields (\ref{approx_ska}) (see (4) and (5) with
$t=t'=1/4$ and $z=4$ in \cite{hanis93}). Thus we obtain
\begin{equation}
\label{finresper}
E^{(2)} < \frac{4\delta}{f}\int\limits_{-1}^{\varepsilon_{\rm F}}
d\varepsilon \frac{A_\pm(\varepsilon)}{\lambda-\varepsilon}
\end{equation}
where $f= \delta(\delta-1) -e_1^2$, $\lambda=e_1/f$, and
$A_\pm$ from (\ref{hamueha}).
The evaluation of the right hand side of (\ref{finresper}) 
yields the dashed curves in fig.\ \ref{fig:tstrich_4}. 
The essence of fig.\ \ref{fig:tstrich_4}
is that the saturated ferromagnetic state is  unstable in the limit 
$t\to 0$ for all fillings. The small region where $E^{(2)}_{\rm FM}$
lies below the upper bound for $E^{(2)}_{\rm AFM}$ does not count 
since we know that at these dopings (and for larger dopings)
already  the pure square lattice  at
$t=0$ has no saturated ferromagnetic ground state, see e.g.\ 
\cite{hanis93,wurth96}.

We wish to draw 
the reader's attention to the fact that the comparison in
fig.\ \ref{fig:tstrich_4} is quite different from the main theme of this paper 
which is based on single spin flip energies.
Here the global stability is tested with  a completely
different, antiferromagnetic state.   We learn from the perturbative argument
that in fig.\ \ref{fig:tstrich_3} the true lines
$n(t')$ comprising the {\em global} Nagaoka stability region have to converge
both to the point $t'=1,n=1$.

\subsection{Simple cubic lattice}

\begin{figure}[htb]
 \setlength{\unitlength}{1cm}
 \begin{picture}(8,6)
   \includegraphics{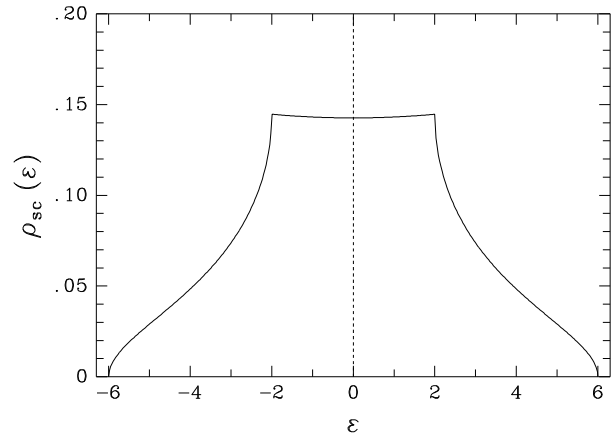}
 \end{picture} \par
 \caption{ \label{fig:sc_1} DOS for the simple cubic lattice
 ($t=1$).}
\end{figure}

The energy dispersion of the simple cubic lattice is 
\begin{equation} \label{sc_dispers}
\varepsilon_{sc} (\underline{k}) 
= -2t \, (\cos k_{x} + \cos k_{y} + \cos k_{z}) \ .
\end{equation}
The calculation of $\rho_{sc} (\varepsilon)$ can be performed by
an integration over the known DOS of the square lattice (see appendix 
\ref{app:dos}).
The maxima and minima of the energy dispersion (\ref{sc_dispers}) are 
$\varepsilon_{\rm t} = z |t|$ and $\varepsilon_{\rm b} 
= -z |t|$, respectively, with the coordination number $z = 6$.
At the band edges the DOS (fig.\ \ref{fig:sc_1}) shows the square root 
 \noindent
behavior which is characteristic for $d=3$.
The van Hove singularities at $\varepsilon = \pm 2t$ correspond to the saddle 
points of the dispersion (\ref{sc_dispers}). 

\begin{figure}[htb]
 \setlength{\unitlength}{1cm}
 \begin{picture}(8,13.8)
   \includegraphics{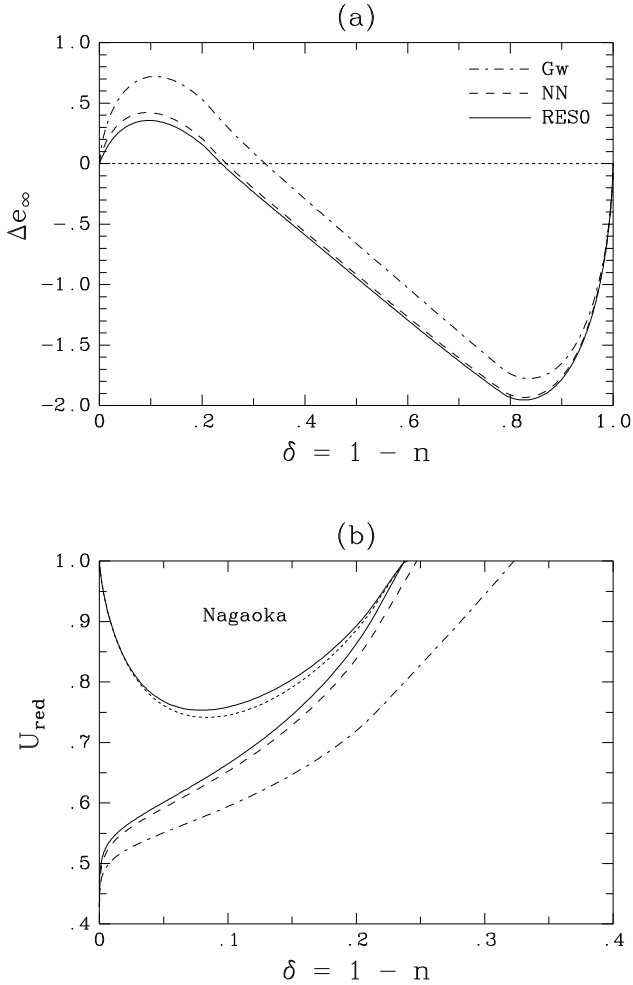}
 \end{picture} \par
 \caption{ \label{fig:sc_2} (a) Spin flip energy at $U=\infty$ as a function of
 the hole density on the \textit{sc} lattice ($t=1$) for Gw, NN, and RES0,
(b) phase diagram ($n<1$): Nagaoka instability lines on the 
    \textit{sc} lattice  for Gw (dashed-dotted), NN (long 
    dashed), RES1 (lower full line), RES2 (short dashed), and RES3 
    (upper full line).} 
\end{figure}

Fig.\ \ref{fig:sc_2}(a) 
shows the spin flip energy at $U=\infty$ for Gw, NN, and RES0. For small hole 
doping the loss of spin-up kinetic energy due to the spin flip is 
sufficiently strong to keep the Nagaoka state stable. With increasing 
$\delta$ the spin flip energy decreases due to the 
gain of kinetic energy for the flipped spin which grows linear with $\delta$
in leading order. The upper bound for the critical
hole density is reduced from $\delta_{\rm cr} = 0.323$ for 
Gw\cite{shast90a,mulle91} to $\delta_{\rm cr} = 0.247$ for NN 
and finally to $\delta_{\rm cr} = 0.237$ for RES0. As in $d=2$, the NN
hopping
term gives the dominant contribution to the decrease of $\delta_{\rm cr}$
while the extension of the spin-up hopping processes to the whole lattice has 
only a small effect.

Roth \cite{roth69} investigated the Nagaoka instability with respect 
to a single spin flip on the \textit{sc} lattice already in 1969, making use
of the so-called two pole approximation instead of the projection method.
It was shown later \cite{tan74,allan82} that the Hilbert subspace considered
in \cite{roth69} is equivalent to the Basile-Elser subspace in the limit
$U \rightarrow \infty$. Roth obtained numerically a critical hole
density of $0.24$ which is consistent with our variational result for RES0.

The phase diagram (fig.\ \ref{fig:sc_2}(b)) for the simple cubic lattice
shows a qualitative difference to the square lattice: The critical $U$ at half
filling obtained for the Gutzwiller single spin flip is not at all improved 
by including NN hopping terms. Even for RES1 
$U_{\rm cr}(\delta=0)$ is still given
by the band width $12 |t|$. This is due to the fact 
that for the \textit{sc} lattice the DOS at the upper band edge 
vanishes while it is nonzero for the square lattice.

As in $d=2$, the ansatz RES2 leads to $U_{\rm cr}
 (\delta = 0) = \infty$ and to a
considerable reduction of the Nagaoka stability regime near half filling.
For the full resolvent ansatz RES3 we finally achieve a minimum critical 
coupling of $U_{\rm cr}^{\rm min} = 48.9 |t|$ (corresponding to $U_{\rm red} = 0.753$) 
below which the Nagaoka state is proven to be unstable for all $\delta$.
\begin{figure}[htb]
 \setlength{\unitlength}{1cm}
 \begin{picture}(8,13.8)
   \includegraphics{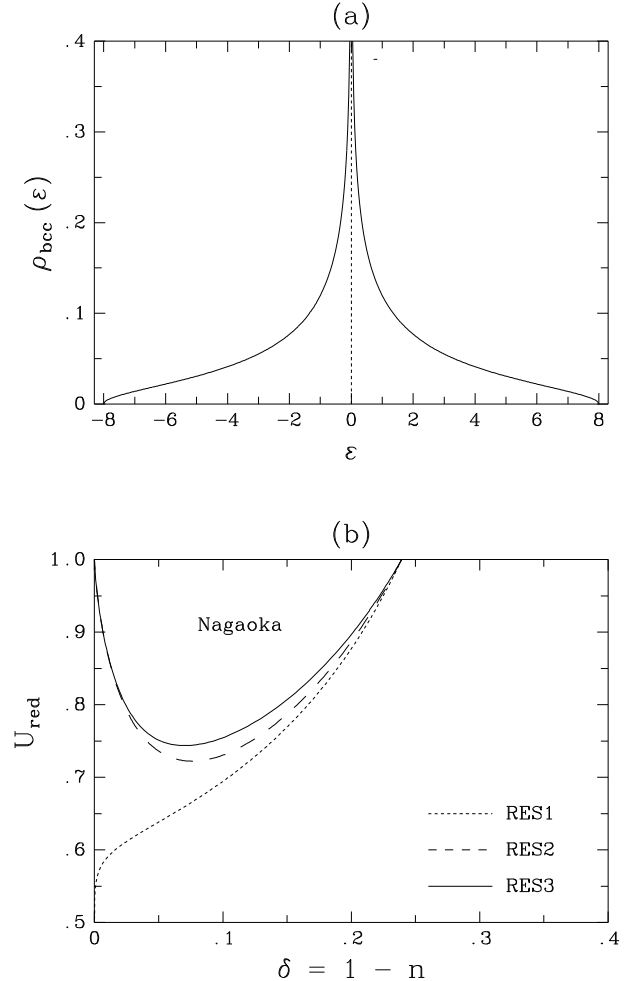}
 \end{picture} \par
 \caption{ \label{fig:bcc} (a) DOS for the \textit{bcc} lattice ($t=1$).
(b) phase diagram ($n<1$): Nagaoka instability lines on the 
    \textit{bcc} lattice  for RES1 (dotted line), RES2 (dashed), and RES3 
    (full line).} 
\end{figure}\noindent
The region left for a possible Nagaoka ground state on the \textit{sc} lattice
is therefore substantially smaller than on the square lattice (RES3 for the
square lattice: $\delta_{\rm cr}= 0.405, U_{\rm cr}^{\rm min} = 36.2|t|$). 
Generally the tendency of the Hubbard model towards a saturated ferromagnetic 
ground state on a $d$-dimensional hypercubic lattice becomes weaker with 
increasing $d$. M\"uller-Hartmann \cite{mulle93b} showed that the critical 
hole density at $U = \infty$ with respect to a Gutzwiller single spin flip 
decreases asymptotically as $\delta_{\rm cr} \propto 1/\sqrt{d \, \ln d}$ 
for $d \gg 1$. In the limiting case of infinite dimensions the ground 
state of the Hubbard model is never fully polarized \cite{fazek90}.

\subsection{bcc lattice}
The \textit{bcc} lattice is another interesting example of a
three-dimensional bipartite lattice. It has a slightly higher
coordination number $z=8$ compared to the simple cubic lattice.
Its dispersion reads
\begin{equation}
\varepsilon_{bcc} (\underline{k}) 
= -8t  \cos( k_{x})\cos( k_{y})  \cos( k_{z}) \ .
\label{bccdispers}
\end{equation}
The calculation of the DOS $\rho_{bcc}(\varepsilon)$ can again be
performed by an integration over the known DOS of the square lattice
(see appendix \ref{app:dos}). The bipartiteness is obvious since
$\varepsilon_{bcc} (\underline{k}+\underline{Q}) +\varepsilon_{bcc} 
(\underline{k}) =0$ with $\underline{Q}=(\pi,\pi,\pi)$. 
For this reason we consider only $n \le 1$.

The DOS is shown in fig.\ \ref{fig:bcc}(a). The square root
singularities  at the band edges are generic for three dimensions.
The least common  feature for a three dimensional lattice is the
squared logarithmic  singularity at zero energy
$\rho_{bcc}(\varepsilon)\approx \ln^2(\varepsilon)/(4\pi^3)$ 
which results from the points  in momentum space where all cosines in
(\ref{bccdispers}) vanish, e.g.  
$\varepsilon_{bcc} (\underline{k})  
\approx -8t (k_x-\pi/2)(k_y-\pi/2)(k_z-\pi/2)$.

Evaluating (\ref{skaband}) for the \textit{bcc} lattice, we find the
critical density  $\delta_{\rm cr}=0.324$ in the Gutzwiller approach.
This is almost the same result as for the simple cubic lattice.
The result $\delta_{\rm cr}=0.239$
for the full ansatz RES0 is also only a tiny bit higher than the RES0 
critical doping for the \textit{sc} lattice.  It appears
that the essential ingredients are indeed the dimensionality and
the bipartiteness as we will see below.

The results for finite interaction are shown in fig.\ \ref{fig:bcc}(b).
The value of $U_{\rm BR}$ is $16.413$.  The reduced interaction values
 are very similar to the ones for the simple cubic lattice.
The ansatz RES1 does not capture the diverging interaction for $n\to 1$
but RES2 yields already the asymptotic behavior of RES3 for $n\to 1$.
The critical interaction is $U_{\rm red, cr}= 0.7438$ for RES3.

As far as the local stability of the Nagaoka state is concerned we
do not find any indication that the \textit{bcc} lattice is more
favorable than the simple cubic lattice. Herrmann and Nolting
\cite{herrm97a,herrm97b} found in the framework of the
spectral density approach an enhanced tendency towards ferromagnetism
for the \textit{bcc} lattice.
They investigated the divergence of the susceptibility in the
paramagnetic phase which is enhanced by the large DOS at zero energy.
Combining their result with ours one might come to the
conclusion that the \textit{bcc} lattice favors a non saturated ferromagnetism
for intermediate coupling and doping.

\subsection{Honeycomb lattice}

Besides the square lattice the honeycomb lattice  (see fig.\ 9 in 
\cite{hanis95}) is another prominent example of a bipartite lattice in $d=2$.
In contrast to the square lattice it is not a Bravais 
lattice, however, but a triangular lattice with a two site basis. The
coordination number is $z=3$ and the band dispersion reads
\begin{equation} \label{honeycomb_dispers}
\varepsilon_{\rm hon} (\underline{k}) = \pm \sqrt{t (3t - 
\varepsilon_{\triangle} (\underline{k}))} \ ,
\end{equation}
where $\varepsilon_{\triangle}(\underline{k})$ stands for the energy 
dispersion  of the triangular lattice to be described in 
(\ref{dreieck_dispers}). Despite this additional complication the 
formulae developed in sect.\ \ref{sect:resapp} via the resolvent method 
hold here as well (see appendix \ref{app:unfrustrated}). 

\begin{figure}[htb]
 \setlength{\unitlength}{1cm}
 \begin{picture}(8,6)
   \includegraphics{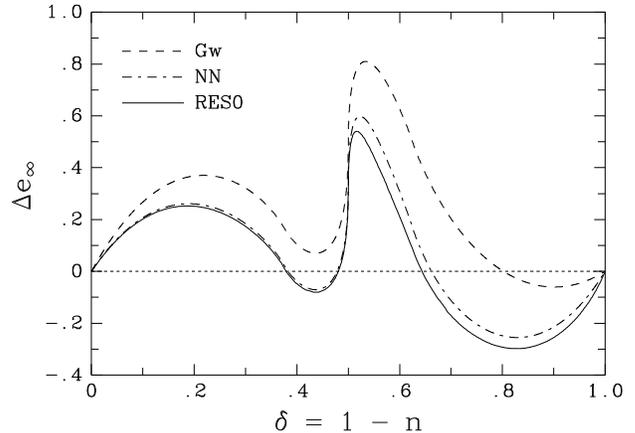}
 \end{picture} \par
 \caption{ \label{fig:honeycomb_1} Spin flip energy at $U=\infty$ as a function
of  the hole density on the honeycomb lattice ($t=1$) for Gw, NN, and RES0.}
\end{figure} \noindent
The instability of the Nagaoka state with respect to Gw and NN
was already discussed in \cite{hanis95}. Here we 
present the improvements obtained by the resolvent method.
The evaluation of RES0 shows that hopping processes 
with a larger distance from the down spin position have only a very small
influence on the optimum spin flip energy at 
$U=\infty$ (fig.\ \ref{fig:honeycomb_1}).

\begin{figure}[htb]
 \setlength{\unitlength}{1cm}
 \begin{picture}(8,6)
   \includegraphics{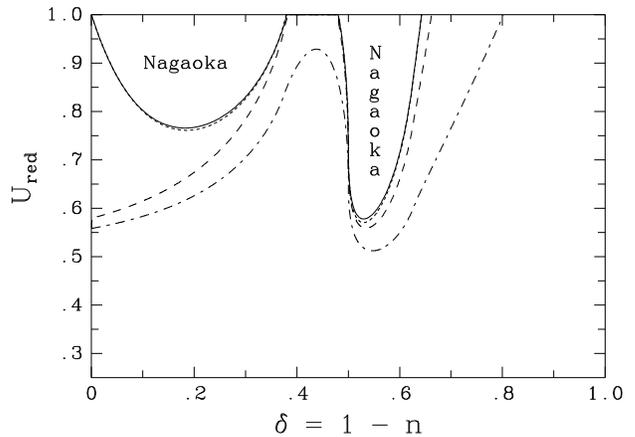}
 \end{picture} \par
 \caption{ \label{fig:honeycomb_2} Phase diagram ($n<1$): 
Nagaoka instability lines on the honeycomb lattice for Gw (dashed-dotted), 
NN (long dashed, almost identical with 
RES1), RES2 (short dashed), and RES3 (full line).}
\end{figure}
The
instability gap ($0.379 \leq \delta \leq 0.481$) between the two 
possible Nagaoka stability regions remains
almost unchanged compared to the result for NN. The upper 
critical hole density is only slightly improved to $\delta_{\rm cr} = 0.643$
from $0.662$ (NN) and $0.802$ (Gw)\cite{hanis95}.

As explained in \cite{hanis95} the Nagaoka stability island in the phase 
diagram around quarter filling (fig.\ \ref{fig:honeycomb_2})
 is mainly due 
to the zero in the DOS at $\varepsilon = 0$, i.e.\ between the two energy 
bands. Since the lattice structure enters the calculation of the optimum
spin flip energy by means of the resolvent method only via the DOS
the stability island is present even for the full resolvent ansatz RES3.
On the other hand, the critical $U$ at half filling diverges for RES2 and 
RES3 and the Nagaoka stability region for small $\delta$   shrinks
compared to the results for NN and RES1. These results and the pronounced
difference between the two minimum values of $U$ ($41.2\, |t|$ for the low 
doping regime and $17.25\, |t|$ for the stability island) corroborate the
previous conjecture\cite{hanis95} that a saturated ferromagnetic ground state 
exists around quarter filling. The lack of a Nagaoka theorem for the honeycomb
lattice\cite{tasak89,hanis95} indicates a degeneracy between the Nagaoka 
state and other possible states near half filling  even at $U=\infty$.

\subsection{Triangular lattice} \label{subs:tri}

The triangular lattice is non-bipartite. It can be decomposed into 
\textit{three} sub-lattices, each of them having triangular structure. 
Investigating the local instability of the Nagaoka state towards a 
Gutzwiller single spin flip a Nagaoka ground state was
excluded on the triangular lattice for less than 
half filling\cite{mulle91,hanis95}. This is in agreement with the Nagaoka
theorem, which predicts a saturated ferromagnetic ground state
at $U=\infty$ only for the half filled lattice \textit{plus} an additional 
electron. Thus we consider henceforth the electron doped case for $t=1$ or,
equivalently, $t=-1$ and $n<1$. 

Each lattice site has $z=6$ nearest neighbors located at the corners of a 
hexagon. The band dispersion is given by
\begin{equation} \label{dreieck_dispers}
\varepsilon_{\triangle} (\underline{k}) =  
2\cos k_x + 4\cos \left(\frac{k_x}{2} \right) \cos \left( 
\frac{\sqrt{3} \, k_y}{2} \right)
\end{equation}
where $\underline{k}$ belongs to the likewise hexagon-shaped first 
Brillouin zone. The upper band edge ($\varepsilon_{\rm t} = 6$) 
is found at the center of the Brillouin zone, whereas the lower band edge 
$\varepsilon_{\rm b} = -3$ is reached at the corners of the 
hexagon. The DOS (see appendix \ref{app:dos} and fig.\ 1 in \cite{hanis95}), 
which can be expressed by a complete elliptic integral, displays a 
logarithmic van Hove singularity at $\varepsilon = -2$.
For $\varepsilon_{\rm F}=-2$
 the Fermi surface forms a hexagon with an area of 3/4 of the 
whole  Brillouin zone. As usual in $d=2$ the DOS at the band edges is 
nonzero ($\rho_b = 4 \rho_t = (\sqrt{3}\pi)^{-1}$).

\begin{figure}[htb]
 \setlength{\unitlength}{1cm}
 \begin{picture}(8,13.8)
   \includegraphics{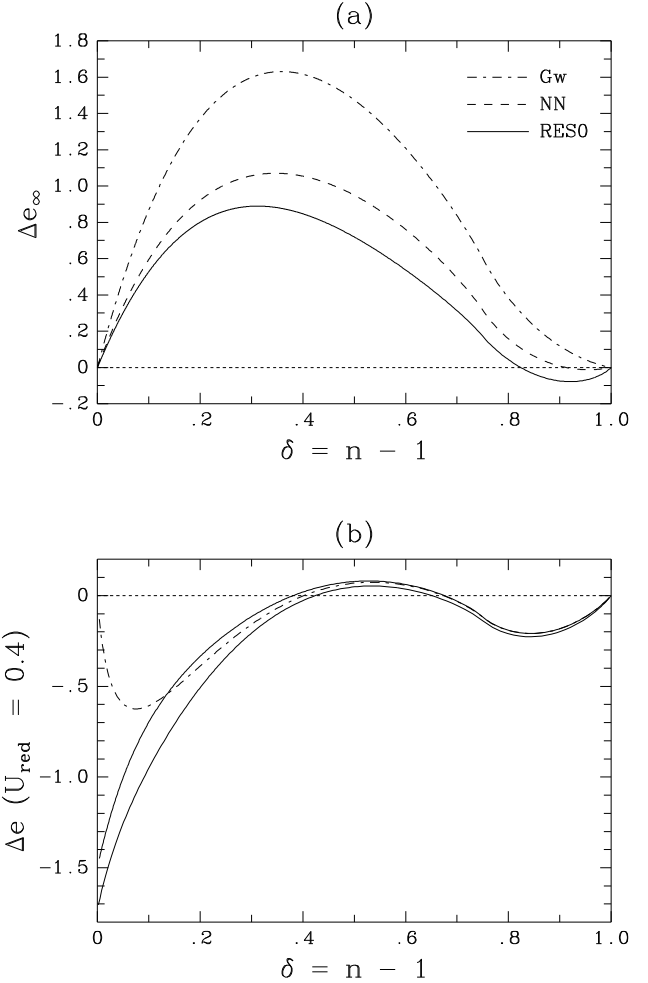}
 \end{picture} \par
 \caption{ \label{fig:dreieck_1} (a) Spin flip energy at $U=\infty$ as 
a function of the hole density on the triangular lattice ($t=-1$) 
for Gw, NN, and RES0, 
(b) spin flip energy at $U_{\rm red} = 0.4$ as a function of 
the hole density on 
the triangular lattice ($t=-1$) for RES1 (dashed-dotted line), RES2
(upper full line), and RES3 (lower full line).}
\end{figure}

\begin{figure}[htb]
 \setlength{\unitlength}{1cm}
 \begin{picture}(8,6)
   \includegraphics{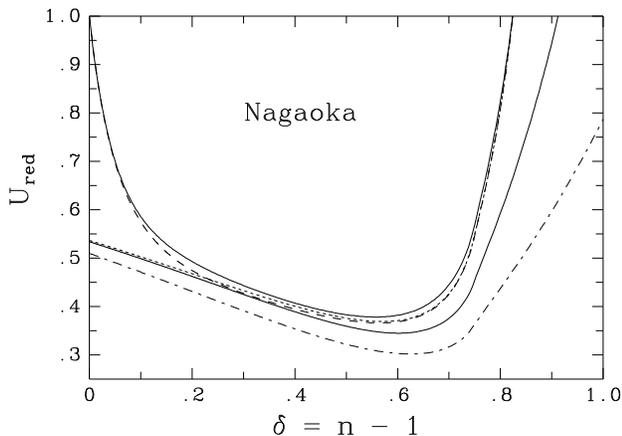}
 \end{picture} \par
 \caption{ \label{fig:dreieck_2} Phase diagram ($n>1$): 
Nagaoka instability lines on the triangular lattice for Gw (dashed-dotted), 
NN (lower full line), RES1 
(short dashed), RES2 (long dashed), and RES3 (upper full line).}
\end{figure}

In contrast to the square lattice, the Nagaoka state remains stable 
towards Gw for all fillings $n>1$
at $U=\infty$\cite{shast90a,mulle91}. The corresponding spin flip energy as a
function of $\delta$ is depicted in fig.\ \ref{fig:dreieck_1}(a). Evaluating
NN, however, a negative spin flip energy is found above 
$\delta_{\rm cr} = 0.912$ proving the instability of the Nagaoka state 
in the low density limit. The resolvent ansatz RES0 lowers the spin flip energy
further and implies $\delta_{\rm cr} = 0.824$ (fig.\ \ref{fig:dreieck_1}(a)).
The difference $\Delta \delta = 0.088$ between the results
obtained for NN and RES0 is eight times larger than the one for  the square 
lattice ($\Delta \delta = 0.011$).
 \noindent
 This  demonstrates the importance
of the spin-up hopping  processes for the instability of Nagaoka ferromagnetism
on the triangular lattice. The reason is that due to the large hole densities 
under consideration, the probability to find unoccupied sites near the 
flipped spin is quite high. The same line of reasoning applies also for
$U<\infty$, see fig.\ \ref{fig:dreieck_2}.
 \noindent

Previously we investigated a variational state which restricts 
the hopping processes to a 31-site cluster around the position of the
flipped spin\cite{hanis95} and obtained $\delta_{\rm cr} = 0.887$.
Although the nearest neighbor processes once again are the most important
ones, the number of relevant hopping processes 
on the triangular lattice turns out to be much larger than on the square 
lattice. Hence the critical hole density $\delta_{\rm cr} = 0.824$
found by the resolvent method is
essentially lower than the one found from the finite cluster calculations.
 Moreover, the evaluation of
RES0 requires much less analytical and numerical effort than the iterative 
extension of the variational ansatz by additional hopping processes.
For details on the application of the resolvent method to the triangular
lattice see appendix \ref{app:tri}.

Near half filling the influence of the majority spin hopping processes 
contained in NN and RES1 (which suppress double occupancies) on the Nagaoka 
stability is negligible, as expected (fig.\ \ref{fig:dreieck_2}). 
In contrast to this the resolvent ansatz RES2 with nearest neighbor 
hopping processes \textit{creating} double occupancies 
leads to a negative spin flip energy near half filling for all 
$U<\infty$ and hence to a divergence of $U_{\rm cr} (n=1)$ 
(fig.\ \ref{fig:dreieck_2}). It turns out, 
however, that for larger hole densities, when the exchange effect looses its
importance,  RES2 is somewhat \textit{less} successful than RES1. 
The plot of the spin flip energy as  a function of $\delta$ for the 
comparatively small on-site repulsion $U_{\rm red} = 0.4$ in fig.\ 
\ref{fig:dreieck_1}(b) demonstrates that above $\delta \simeq 0.12$ 
the creation of extra holes near the flipped spin as described by RES2 is 
energetically unfavorable. The full resolvent ansatz RES3, comprising RES1 
and RES2, gives of course the best lower bound for the Nagaoka instability 
line $U_{\rm cr} (\delta)$. The minimum critical coupling obtained for RES3 
is $U_{\rm cr}^{\rm min} = 9.62 |t|$ ($U_{\rm red} = 0.378$),
 the critical hole density
at $U=\infty$ is given by the RES0 value $\delta_{\rm cr} = 0.824$. Hence the 
region for a possible Nagaoka ground state on the triangular lattice 
appears to be much larger than on the bipartite square and honeycomb 
lattices.

\subsection{Kagome lattice}

\begin{figure}[htb]
 \setlength{\unitlength}{1cm}
 \begin{picture}(8,6)
   \includegraphics{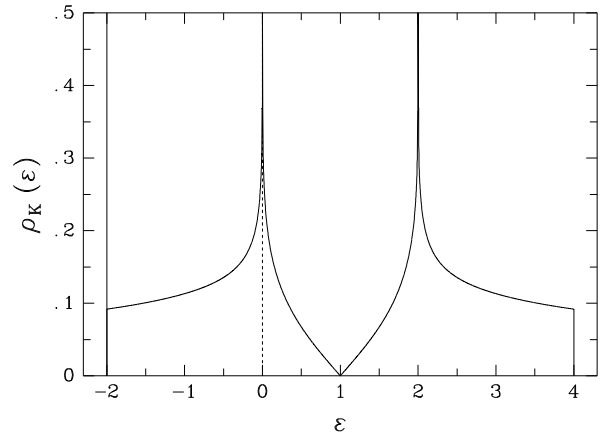}
 \end{picture} \par
 \caption{ \label{fig:kagome_1} DOS for the kagome lattice
($t=-1$).}
\end{figure}

Taking the kagome lattice as an example of a frustrated non-Bravais lattice 
we want to demonstrate that the resolvent method works also for this class
of lattices. Representing the line graph \cite{anmerklinegraph} of the 
honeycomb lattice the kagome lattice (for $t<0$) shows a flat, i.\ e.\ 
dispersionless band with spectral weight 1/3 at the lower band edge 
$\varepsilon_{\rm b} = -2 |t|$ (fig.\ \ref{fig:kagome_1}).
 \noindent
 All line graphs display such a flat band\cite{mielk91a}.
 The kagome lattice is 
the first and the most prominent example of so-called flat-band 
ferromagnetism\cite{mielk92,tasak95}. 
A macroscopic degeneracy of the lowest single 
particle energy leads for certain band fillings to a unique saturated 
ferromagnetic ground state. Mielke \cite{mielk91a}
 proved that the Nagaoka state
is the unique ground state of the Hubbard model on the kagome lattice 
for all $U>0$ at $n = 1/3$.
Although in the flat-band regime every ground state of the
Hamiltonian (\ref{def:model}) is a simultaneous eigenstate of $H_{\rm kin}$ and
$H_{\rm pot}$, the uniqueness of the ground state is not trivial.
For $n <  1/3$ the fully polarized ground state is not unique
\cite{mielk93b}.

The kagome lattice can be considered as a triangular lattice with a basis
of three lattice points\cite{hanis95}, see also appendix
\ref{app:kagome}. Besides the flat band 
$\varepsilon (\underline{k}) = 2t$ the diagonalization of 
$H_{\rm kin}$ leads to the two dispersive bands
\begin{equation}
\label{kagdispers}
\varepsilon_{K} (\underline{k}) = -t \left( 1 \pm \sqrt{3 - 
\varepsilon_{\triangle} (\underline{k})/t} \right) 
\end{equation}
where $\varepsilon_{\triangle} (\underline{k})$ stands for the dispersion
(\ref{dreieck_dispers}) of the triangular lattice. 
For the kagome lattice the resolvent method requires
less effort than for the triangular lattice with $t<0$ since the lower 
band edge $\varepsilon_{\rm b} = -2 |t|$ is reached at 
$\underline{k}_{\rm b} =  \underline{0}$ for one
dispersive band and for the flat band of course. Thus  $\underline{q} =
 \underline{0}$ is the optimum momentum as for bipartite lattices.
Hence all
lattice dependent quantities appearing in our formulae can be calculated
as integrals over the  DOS $\rho_{K} (\varepsilon)$.
 For $t=-1$ one finds the DOS of the kagome
lattice (see fig.\ \ref{fig:kagome_1} and appendix \ref{app:dos}) as
\begin{equation}
\rho_{K} (\varepsilon) = \frac{1}{3} \, \delta(\varepsilon +2) +
\frac{2}{3} | \varepsilon - 1| \cdot \rho_{\triangle} \left(
(\varepsilon -1)^2-3 \right); 
\end{equation}
$\rho_{\triangle} (\varepsilon)$ is the DOS
of the triangular lattice. Nevertheless, the fact that 
the kagome lattice is \textit{not} a Bravais lattice induces
some changes in the analytic expressions for the spin flip energy
(appendix \ref{app:kagome}).

\begin{figure}[htb]
 \setlength{\unitlength}{1cm}
 \begin{picture}(8,13.8)
   \includegraphics{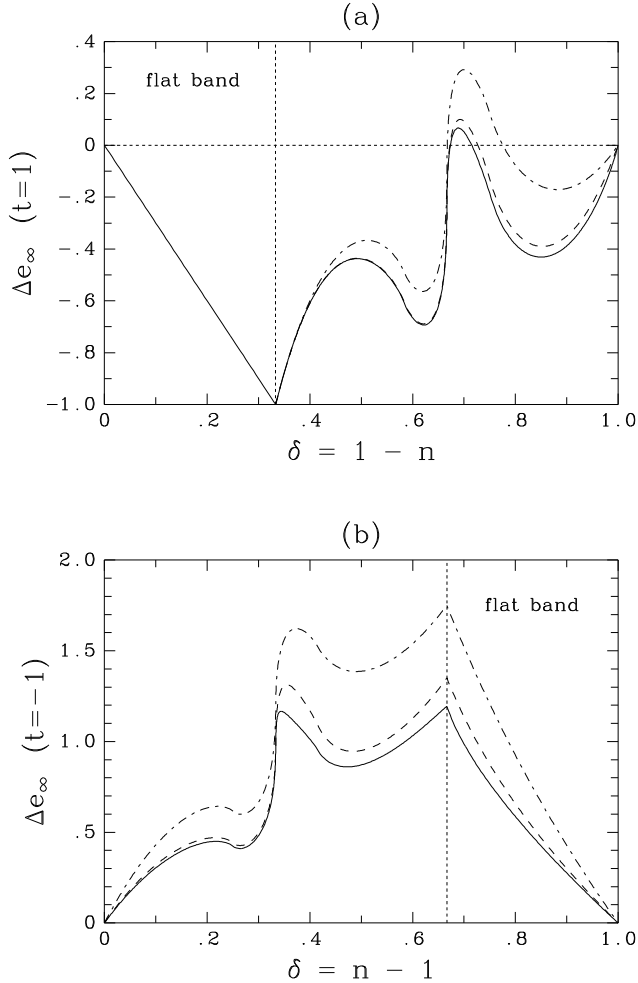}
 \end{picture} \par
 \caption{ \label{fig:kagome_2} Spin flip energy at $U=\infty$ as a function
of the hole density on the kagome lattice for Gw (dashed-dotted), NN (dashed), 
and RES0 (a) for $t=1$,  (b) for $t=-1$.}
\end{figure}
\begin{figure}[htb]
 \setlength{\unitlength}{1cm}
 \begin{picture}(8,6)
   \includegraphics{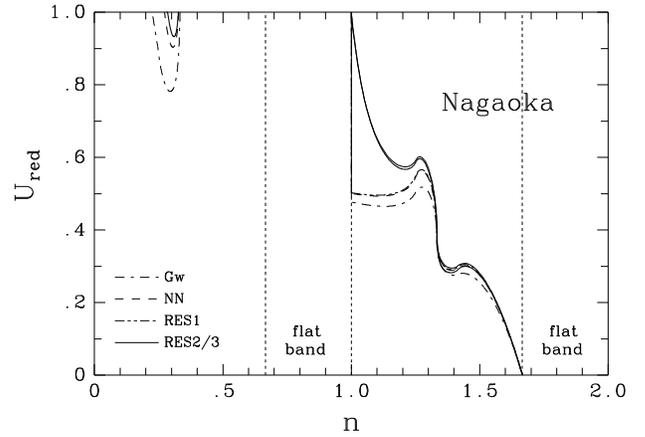}
 \end{picture} \par
 \caption{ \label{fig:kagome_3} Phase diagram: Nagaoka instability lines on the
kagome lattice for Gw, NN, RES1 (lower full 
line), RES2, and RES3 (upper full line). For $n<1$ the
difference between RES1, RES2, and RES3 is less than the line width.}
\end{figure}

Fig.\ \ref{fig:kagome_2}(b)
 shows the spin flip energy for $U=\infty$ and
$t=-1$ as a function of $\delta$  for RES0 compared to the Gw and NN 
results obtained in \cite{hanis95}. As for the honeycomb lattice the 
enhancement of the Nagaoka stability for $\delta \rightarrow 1/3$ is due to 
the zero in the DOS. The effect of the additional spin-up hopping processes 
contained in RES0 is most pronounced for $\delta > 1/3$. But the spin flip 
energy remains positive for all band fillings. Note that the exact result 
in the flat-band regime is a zero spin flip energy\cite{mielk93b,hanis95}.
The phase diagram for $n>1$ in fig.\ \ref{fig:kagome_3}
 \noindent
 shows a
strong tendency towards Nagaoka ferromagnetism also beyond the flat-band 
regime, where we find the Nagaoka state to be stable for all $U>0$. 
There is only a marginal difference between the Nagaoka instability lines 
for NN and for RES1, since the values of $U$  under consideration are too 
small to allow a significant reduction of the spin flip energy 
by Basile-Elser hopping processes. Near half filling, however, we are able to 
restrict the Nagaoka stability region by RES2, i.e.\ by taking  
antiferromagnetic exchange processes into account. 
As for the triangular lattice, for a certain range of filling around 
$n = 3/2$ away from half filling RES2 gives a weaker bound for 
$U_{\rm cr} (\delta)$ than RES1.

For positive hopping matrix element $t$ the flat band is found at the 
\textit{upper} band edge. The flat-band regime for $n<1$  corresponds to 
hole densities $0 \leq \delta \leq 1/3$. Since 
$\underline{k}_{\rm b} = \underline{0}$ \textit{and} 
$\varepsilon_{\rm b} = -zt$ the resolvent method formulae
are those of the bipartite lattices (see appendix \ref{app:unfrustrated}).
Figs.\ \ref{fig:kagome_2}(a) and \ref{fig:kagome_3} show that the small 
Nagaoka stability island found previously\cite{hanis95} for very large $U$ 
around quarter filling is still present for RES0 -- RES3. The upper 
critical hole density is reduced from 0.727 (NN) to 0.715 (RES0) and 
$U_{\rm cr}^{\rm min}$ reaches $191.5 |t|$ (RES3) 
instead of $129.5 |t|$ (NN) though.
These results may indicate that this stability island really provides an 
example of a saturated ferromagnetic ground state on a non-bipartite lattice 
for less than half filling. Its origin\cite{hanis95} is the zero in the DOS 
enhancing the stability of the Nagaoka state around $\delta = 2/3$.

\subsection{fcc and hcp lattices}
The \textit{fcc} and \textit{hcp} lattices as the most prominent 
close-packed lattices in $d=3$  are found in numerous real substances among 
them the ferromagnetic transition metals Ni (\textit{fcc}) and 
Co (\textit{hcp}). The face-centered cubic lattice is
a Bravais lattice with coordination number $z=12$.  Its band dispersion
\begin{eqnarray} \nonumber
&&\varepsilon_{fcc} (\underline{k}) =\\
\label{fcc_dispers}
&& \quad -4t \, (\cos k_x \cos k_y + \cos k_x
\cos k_z + \cos k_y \cos k_z)
\end{eqnarray}
is related to the dispersion (\ref{sc_dispers}) of the simple cubic 
(\textit{sc}) lattice via\cite{mulle91}
\begin{equation} \label{ident:sc_fcc}
\varepsilon_{fcc} (\underline{k}) = 
- \frac{\varepsilon_{sc}^2 (\underline{k})}{2t} 
-\frac{\varepsilon_{sc} (2\underline{k})}{2} + 3t \ .
\end{equation} 
The hexagonal close-packed lattice (also with $z = 12$) is not a Bravais 
lattice but a hexagonal lattice with basis.
\begin{figure}[htb]
 \setlength{\unitlength}{1cm}
 \begin{picture}(8,13.8)
   \includegraphics{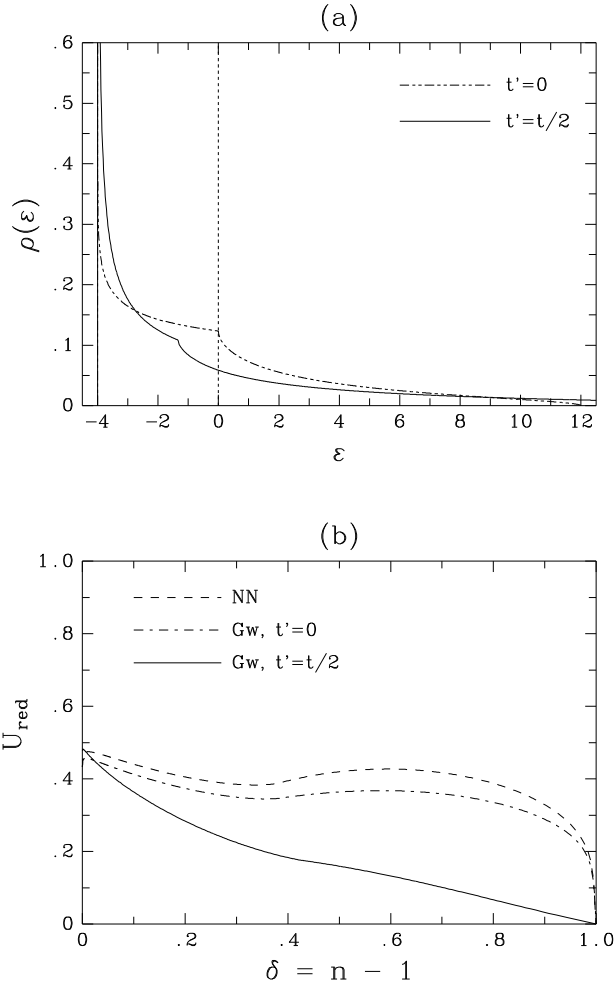}
 \end{picture} \par
 \caption{ \label{fig:fcc} (a) Identical DOS for the \textit{fcc} and 
\textit{hcp} lattices ($t=-1$) and DOS for the \textit{fcc} lattice with 
additional next nearest neighbor hopping ($t' = t/2$),
(b) phase diagram ($n>1$): Nagaoka instability lines on the \textit{fcc} 
lattice for Gw and NN and on the $t$-$t'$-\textit{fcc} lattice with $t'=t/2$ 
for Gw.}
\end{figure} \noindent
 Within the hexagonal planes, 
which we assume to be parallel to the $xy$ plane,
the $\underline{k}$-dependence of the two energy bands reduces 
to the energy dispersion (\ref{dreieck_dispers}) of the triangular 
lattice\cite{mulle91}:
\begin{eqnarray} \nonumber
&&\varepsilon_{hcp} (\underline{k}) = \varepsilon_{\triangle} (k_x, k_y) \\
&& \label{dispers_hcp} \qquad \pm
2t \cos \left(\sqrt{2/3}\, k_z \right) \cdot \sqrt{3 - \varepsilon_{\triangle}
(k_x, k_y)/t} \ .
\end{eqnarray}
Hence it is possible to compute the DOS of the \textit{hcp} lattice by 
integration over $\rho_{\triangle} (\varepsilon)$ (see appendix \ref{app:dos}).

Modelling the \textit{fcc} and \textit{hcp} structures with close-packed 
spheres, the sequence of layers with different positions of the sphere 
centers is known to be ABABAB... for the \textit{hcp} 
and ABCABC... for the \textit{fcc} lattice. Diagonalizing the kinetic part 
of the  Hamiltonian (\ref{def:model}) in each of the hexagonal 
planes it turns out that 
the terms reflecting the different arrangement of the planes disappear if 
one chooses the Fourier transformation in a convenient way. The 
\textit{densities of states} for the \textit{fcc} and for the \textit{hcp} 
lattices are  therefore \textit{identical}
as well as our variational results on the stability of the Nagaoka state 
with respect to Gw and NN.

For less than half filling a saturated ferromagnetic ground state was
already excluded due to the complete instability of the Nagaoka state towards
Gw at $U=\infty$\cite{mulle91}. Therefore we only
investigate the case of more than half filling which corresponds to $t<0$.  
The DOS (fig.\ \ref{fig:fcc}(a))
displays  the $d=3$ square root behavior 
$\rho (\varepsilon) \propto \sqrt{\varepsilon_{\rm t} - \varepsilon}$ 
at the upper band edge $\varepsilon_{\rm t} = z |t|$. For the \textit{fcc}
lattice the lower 
band edge $\varepsilon_{\rm b} = -4|t|$  is reached on different 
\textit{lines} in $\underline{k}$-space which intersect in several critical 
points located on the border of the Brillouin zone. This reduces the 
``effective dimensionality'' of the van Hove singularity by one 
and leads to a \textit{logarithmic singularity} in the DOS.
As a consequence, the Nagaoka state remains stable in the low density limit 
(corresponding to $n \rightarrow 2$ for $t>0$) for all $U>0$ with respect to 
Gw and NN (fig.\  \ref{fig:fcc}(b)). This result indicates the strong 
tendency towards 
Nagaoka ferromagnetism on the \textit{fcc} lattice, especially in comparison 
with the triangular lattice where we proved the \textit{instability} of the
Nagaoka state even for $U=\infty$ in the low density  limit. 
Also for low and intermediate hole doping the extension of the Gutzwiller 
wave function by nearest neighbor hopping processes yields only a slight
reduction of the Nagaoka stability region in the phase diagram.  This is in 
sharp contrast, for example, to the situation on the \textit{sc} lattice.
The resolvent method was not applied to the \textit{fcc} and \textit{hcp}
lattices since three dimensional momentum integrals would have to be performed
in order to calculate $h(\omega)$ and $\overline{h} (\omega)$ (see 
appendix \ref{app:tri}).  

As for the square lattice (see section \ref{subs:tprime}) the
particle-hole asymmetry of the 
DOS is even enhanced if one extends $H_{\rm kin}$ by electron 
hopping between \textit{next nearest neighbor sites} 
with a hopping amplitude $t'$. 
On the \textit{fcc} lattice, these sites form a simple cubic
structure such that the additional contribution to the dispersion exactly 
compensates the second term on the right hand side of (\ref{ident:sc_fcc}) 
if $t' = t/2$. In this case the DOS for the $t$-$t'$-\textit{fcc} lattice 
is connected to the DOS of the \textit{sc} lattice  via 
\begin{equation} \label{ident:ttstrich}
\rho_{t-t'} (\varepsilon) = \sqrt{\frac{2}{3(1-\frac{\varepsilon}{3t})}} \
\cdot \ \rho_{sc} \left(\sqrt{6(1 - \frac{\varepsilon}{3t})t} \right) 
\end{equation}
and therefore finally simplifies to an integral over 
$\rho_{\Box} (\varepsilon)$ (see appendix \ref{app:dos}). The next nearest 
neighbor hopping with amplitude $t' = t/2$ creates a \textit{square root} 
divergence of the DOS at the lower band edge (for $t,t' <0$) 
in contrast to the \textit{logarithmic} singularity obtained for $t'=0$ 
(see fig.\ \ref{fig:fcc}(a)). 

The Nagaoka instability line for a Gutzwiller single spin flip
on the \textit{fcc} lattice with $t' = t/2$ is compared 
with the result for the simple \textit{fcc} lattice ($t'=0$) 
in fig.\ \ref{fig:fcc}(b). The more pronounced singularity of the DOS at the 
lower band edge leads to an even more pronounced stability of the Nagaoka 
state in the low density limit. We find $U_{\rm cr} (\delta) 
\propto 1 - \delta$ 
instead of $U_{\rm cr} (\delta) \propto 1/\log(1-\delta)$ for $t' =0$. 
The slight increase of the critical $U$ at half filling is due to the
different band width of the $t$-$t'$-$U$ model ($18|t|$ instead of $16|t|$).

\section{Conclusions} \label{sect:conclus}

In summary, we investigated the stability of the Nagaoka state for
a series of  two- and three-dimensional lattices:
 the square $t$-, the square $t$-$t'$-, the simple cubic, the \textit{bcc},
the honeycomb, 
the triangular, the kagome, and the \textit{fcc} (\textit{hcp}) lattice.
The results were mostly variational in nature and concerned the energy
change due to a single spin flip. By the resolvent approach
the eigenvalue problem in the
variational subspace was reduced  to a matrix
inversion problem. The relatively simple structure of the matrices
under consideration permits to convert the matrix inversion into a 
scalar inversion (or the inversion of a $2\times2$ matrix).
For the $t$-$t'$ square lattice a perturbative approach in $t$ was
used as well for investigating the $t/t'\to 0$ limit.

The ansatzes RES0 - RES3 are particularly  simple for
 unfrustrated,
isotropic, homogeneous lattices with nearest neighbor hopping
(sect.\ \ref{sect:resapp}, appendix \ref{app:unfrustrated}).
For frustrated, non-bipartite lattices our approach is still
tractable, though more cumbersome. To demonstrate its
tractability we derived formulae for the triangular lattice
(non-bipartite, Bravais lattice) and for the kagome lattice
(non-bipartite, non-Bravais lattice).

We believe that our variational criteria are well suited in order to 
investigate the lattice dependence of saturated ferromagnetism in the Hubbard 
model since they cover the most relevant local excitations in the spin-up
Fermi sea but are still simple enough to be evaluated routinely
on various lattices 
in various dimensions. Local and non-local band narrowing effects
are present in these approaches. Since this fact is not obvious in the
complete approaches we resort to the previous result (\ref{skaband}).
The factor $\delta(1-(e_1/\delta z t)^2)$ clearly describes the band
narrowing of the flipped spin. It comprises two factors one of which
is local ($\delta$) and thus survives also in the limit $d\to\infty$.
The other factor ($1-(e_1/\delta z t)^2)$ is very important as well since it
vanishes equally on $\delta\to0 $. But in the limit $z\to \infty$ on
scaling $t\propto 1/\sqrt{z}$ \cite{metzn89a} the latter factor
degenerates to unity. This clearly shows its non-local character.
 Note the importance of the sequence of limits.
The more sophisticated variational approaches discussed in the present
work (RES0 - RES3) comprise the ansatz (\ref{ansatz:gw}). Thus they
contain also local and non-local band narrowing effects.
The other main effect is a direct energy lifting of the minority
electron due to the infinite (or large) on-site repulsion. Since
the minority electron blocks a site, the majority electrons loose the
kinetic energy related to hopping onto or from this site, namely
$e_1$. This is seen best in the kinetic matrix elements 
in (\ref{xelemb}) or in the energy denominator (\ref{fdef}).
Futhermore, we like to draw the reader's attention to the 
non-orthogonality as it can be discerned in (\ref{xelema}).
It is very difficult to comprehend its effect intuitively.
But we know from the extensive efforts to reduce the critical
doping by including more and more correlations \cite{wurth96}
that this non-orthogonality hinders the spin flip to gain enough
energy to destabilize the Nagaoka state. The added states do not reduce
the critical doping any further since they do not really enhance the
accessible Hilbert space.

Besides the achievement of easily evaluated ansatzes the comparison
of the phase diagrams presented here yields the following  main results.
For bipartite lattices the possible Nagaoka region shrinks
rapidly with increasing coordination number $z$ (cf. square
and simple cubic lattice). Herrmann and Nolting did not 
investigate low-dimensional lattices because they suppose that
ferromagnetism is excluded in $d=1$ and 2 by the 
Mermin-Wagner theorem \cite{herrm97a,herrm97b}.
Note, however, that neither the  Mermin-Wagner theorem
 makes any statement on ground states \cite{mermi66}
nor any extended theorem can exclude
a ferromagnetic ground state since the total spin as \textit{conserved}
quantity is not affected by quantum fluctuations.
The shrinking of the Nagaoka region on increasing coordination number
 can be understood from the lowering of the DOS at the band edges or,
equivalently, as effect of a lower and lower band edge $\varepsilon_{\rm b}$.

For the above reasons we investigated low-dimensional non-bipartite lattices
where low DOS at the lower band edge can be avoided. Indeed, we found that the possible Nagaoka regions are enlarged
considerably. This is true for electron doping for conventional hopping
($t>0, n>1$) whereas saturated ferromagnetism in the hole doped region
($t>0, n<1$) can be excluded by our results. 
Treating the electron doping also as hole doping after a particle-hole
transformation, i.e.\ $t>0, n>1$ $\to$ $t<0, n<1$, this phenomenon is
easily understood: $\varepsilon_{\rm b}(t<0) < \varepsilon_{\rm b}(t>0)$.
The ratio of the lower band edges is $2$ for the triangular and the
kagome lattice, and $3$ for the \textit{fcc} and \textit{hcp} lattice. In infinite
dimensions it becomes even $\infty$ for the generalizations of the \textit{fcc} lattice
\cite{mulle91,uhrig96a,ulmke96}. For these generalizations
one has
$\varepsilon_{\rm b}(t>0) / \varepsilon_{\rm b}(t<0) \propto \sqrt{d}$.

The above observations concern already the asymmetry of the density of
states. Our results clearly show that a large asymmetry favors ferromagnetism.
It is most useful to have a {\em large DOS at the lower band edge} in the hole
doping picture to stabilize the Nagaoka state. Note that this is {\em not}
equivalent to the well-known Stoner criterion $U\rho(\varepsilon_{\rm F})>1$
which concerns only the DOS at the Fermi level \cite{vollh97}.
 The best situation is to
have a strongly diverging singularity at the lower band edge or close to it
as we found
in the investigation of the $t$-$t'$ model with  tunable DOS and as was also
observed previously for \textit{fcc}-type lattices 
\cite{uhrig96a,ulmke96,herrm97b}.

Our results concerning the $t$-$t'$ model extend previous 
ones \cite{hlubi97} since we treat all ratios of $t$ and
 $t'$ and all fillings $n$. Hlubina
{\it et al}. focussed on the Fermi levels at the van-Hove singularity. Thus
the Stoner criterion is  at the basis of their investigation albeit it
goes technically beyond this mean-field criterion.

Herrmann and Nolting used a two-pole method (SDA: self-consistent spectral
density approach)
to investigate ferromagnetism
for the simple cubic, the $d=\infty$ hypercubic,
 the $d=\infty$ \textit{fcc}, and the \textit{bcc}
 lattice \cite{herrm97a,herrm97b}
at zero and at finite temperature.
Their qualitative findings for zero temperature are similar to ours.
We like, however, to point out that the two-pole method they employ
is indeed a generalization of the Gutzwiller ansatz in (\ref{ansatz:gw})
with $f=0$
to finite temperatures and non-saturated magnetizations.
 For $T=0$ and saturation it reduces to (\ref{ansatz:gw}) with $f=0$.
Thus it is not astounding that they found a good agreement to the results
of Shastry {\it et al.}\ \cite{shast90a}. Our approaches go far beyond 
(\ref{ansatz:gw}) (barring the question of the
 extendability to finite temperatures). This can be seen for instance
 for the simple cubic lattice where we found $\delta_{\rm cr}=0.237$ 
well below  $\delta_{\rm cr}=0.32$ \cite{shast90a,herrm97b}. Already
Roth found by numerical calculation in the variational subspace of RES0
the number $\delta_{\rm cr}=0.24$ \cite{roth69}.
For the \textit{bcc} lattice one finds again that the SDA method
 \cite{herrm97b}
reproduces the Gutzwiller result $\delta_{\rm cr}=0.324$ for 
saturated ferromagnetism whereas RES0 yields a considerably lower
value of $\delta_{\rm cr}=0.239$.
Thus one is led to the conclusion that the SDA two-pole method cannot
be exact as claimed  in the strong coupling limit \cite{herrm97a}.

\section*{Acknowledgements:}
This work was performed within the research program of the
Sonderforschungsbereich 341 supported by the Deutsche Forschungsgemeinschaft.
The authors gratefully acknowledge useful discussions with Peter Wurth and
Burkhard Kleine.


\begin{thebibliography}{10}
\bibitem{anmerklinegraph}
The line graph of a graph (lattice) is constructed by converting
edges into vertices. Two vertices of the line graph are linked
if the corresponding edges on the original graph had a vertex in common.
\end{thebibliography}

\appendix
\section{The full resolvent ansatz RES3} \label{app:res3}
Computing the elements of the matrix $\omega {\bf P} - {\bf L}$ (see
(\ref{invert})) using Wick's theorem we obtain
\begin{mathletters}
\begin{eqnarray}
&&({\bf P_1})_{\underline{k}_1 \underline{k}_2} = n \cdot 
\delta_{\underline{k}_1 \underline{k}_2} + |\Lambda|^{-1}\ , 
\\
&& ({\bf D_1})_{\underline{k}_1 \underline{k}_2} = \nonumber
  [n (\omega - \varepsilon(\underline{k}_2))\ , \\
&&
+ e_{1} (1+ \varepsilon
(\underline{k}_2 - \underline{k}_{\rm\scriptstyle b})/(zt))] \cdot
\delta_{\underline{k}_1 \underline{k}_2} + |\Lambda|^{-1} (\omega - 
\varepsilon_{\rm\scriptstyle b})
\\
&& {\bf N}_{\underline{k}_1 \underline{k}_2} =\nonumber \\
&&
  - |\Lambda|^{-1} (\varepsilon (\underline{k}_1) - 
\varepsilon (\underline{k}_2)- \varepsilon (\underline{k}_1-\underline{k}_2 
-\underline{k}_{\rm\scriptstyle b}) + \varepsilon_{\rm\scriptstyle b})).
\end{eqnarray}
\end{mathletters}
Since only the positions of the creation and the annihilation operator are
interchanged between the states $|\Phi_{\underline{k}} \rangle$ and
$|\Psi_{\underline{k}} \rangle$ one gets ${\bf P}_2$ and ${\bf D}_2$ from
${\bf P}_1$ and ${\bf D}_1$ substituting $n$ by $\delta$, $\omega$ by
$\omega -U$ and $\varepsilon(\underline{k})$ by $-\varepsilon(\underline{k})$.
As for ${\bf D}_1$ (\ref{invert}), 
${\bf D}_2$ contains a diagonal matrix and a $\underline{k}$-independent part:
\begin{equation} \label{part1}
{\bf D_2} = {\bf d}_{\bf 2}^{-1} + (\omega - U - 
\varepsilon_{\rm\scriptstyle b}) \, \underline{u} \underline{u}^{+}
\end{equation}
with
\begin{eqnarray}
&&({\bf d}_{\bf 2}^{-1})_{\underline{k}_1 \underline{k}_2} =
\delta_{\underline{k}_1 \underline{k}_2} \cdot [\delta (\omega - U 
+ \varepsilon(\underline{k}_2)) \nonumber \\
&& \quad + e_{1}(1 - \varepsilon(\underline{k}_2 -
\underline{k}_{\rm\scriptstyle b})/(zt))] \ , \ 
(\underline{u})_{\underline{k}} = |\Lambda|^{-1/2}
\end{eqnarray}
for $\underline{k}, \underline{k}_1, \underline{k}_2 \in {\rm FS}$. 
${\bf D}_{\bf 1}^{-1}$ is known already from RES0 (\ref{res0}) 
and we obtain the 
matrix elements of ${\bf N}^{+} {\bf D}_{\bf 1}^{-1} {\bf N}$ for 
$\underline{k}_1, \underline{k}_2 \in {\rm FS}$, $\underline{q}_1, 
\underline{q}_2 \in {\rm BZ} \setminus {\rm FS}$ as
\begin{eqnarray}
&&({\bf N}^{+} {\bf D}_{\bf 1}^{-1} {\bf N})_{\underline{k}_1 \underline{k}_2}
= |\Lambda|^{-2} \sum_{\underline{q}_1 \underline{q}_2}
\Bigg(f(\underline{q}_1) \, \delta_{\underline{q}_1 \underline{q}_2}
\nonumber \\
&& \qquad - 
\frac{\omega - \varepsilon_{\rm\scriptstyle b}}{1 + (\omega - 
\varepsilon_{\rm\scriptstyle b}) h(\omega)} \cdot \frac{
f(\underline{q}_1) f(\underline{q}_2)}{|\Lambda|} \Bigg) \times \nonumber
\\  
&& (\varepsilon_{\rm\scriptstyle b} - \varepsilon(\underline{k}_1 - 
\underline{q}_1 -\underline{k}_{\rm\scriptstyle b}) + 
\varepsilon(\underline{k}_1) - \varepsilon(\underline{q}_1))\times \nonumber
\\
&& (\varepsilon_{\rm\scriptstyle b} - \varepsilon(\underline{k}_2 - 
\underline{q}_2 -\underline{k}_{\rm\scriptstyle b}) + 
\varepsilon(\underline{k}_2) - \varepsilon(\underline{q}_2)).
\label{factor1}
\end{eqnarray}

A remarkable simplification occurs if terms like $\varepsilon(\underline{k}
 - \underline{q})$ factorize to $-\varepsilon(\underline{k})
\varepsilon(\underline{q})/(zt)$. This happens if, as for
hypercubic lattices, every component gives the same contribution to the sum
over $\underline{q}_i$ due to the symmetry of the Brillouin zone.
In this case the corresponding matrix element of the one particle Green's 
function is invariant under permutation of the components. Of course 
the bound state we are looking for has to display
 the same permutation symmetry.
Making use of this argument and assuming $\underline{k}_{\rm\scriptstyle b}
= \underline{0}$, $\varepsilon_{\rm\scriptstyle b} = -zt$ the product 
in the second line of (\ref{factor1}) simplifies to
\begin{equation}
(zt)^{2} \big( \frac{\varepsilon (\underline{k}_1)}{zt} - 1
\big) \big( \frac{\varepsilon (\underline{k}_2)}{zt} - 1
\big) \big( \frac{\varepsilon (\underline{q}_1)}{zt} + 1
\big) \big( \frac{\varepsilon (\underline{q}_2)}{zt} + 1
\big) \ .
\end{equation}
Carrying out the summation over $\underline{q}_{1}$ and $\underline{q}_{2}$ 
 we obtain
\begin{equation} \label{part2}
({\bf N}^{+} {\bf D}_{\bf 1}^{-1} {\bf N})_{\underline{k}_1 \underline{k}_2}
= |\Lambda|^{-1} 
\left(\frac{\varepsilon(\underline{k}_1)}{zt} -1 \right) 
\left(\frac{\varepsilon(\underline{k}_2)}{zt} -1 \right) \alpha
\end{equation}
with 
\begin{eqnarray} \nonumber
\alpha &:=& \gamma^{-1} \Big(e_1 + 2 \delta zt -\delta\Omega +(\Omega+zt)^2
G(\Omega) \\
\label{def:alpha}
&-& \frac{(\Omega+zt)[(\Omega+zt)G(\Omega) - \delta]^2}{n +
(\Omega+zt) G(\Omega)} \Big) 
\end{eqnarray}
depending only on $\omega$ and $\varepsilon_{\rm\scriptstyle F}$, but not
on the indices $\underline{k}_1$ and $\underline{k}_2$.
We define the vector $\underline{w}$ by 
$(\underline{w})_{\underline{k}} := |\Lambda|^{-1/2} \varepsilon
(\underline{k})/(zt)$ for $\underline{k} \in {\rm FS}$. Making use of this
 definition, (\ref{B2def}),
(\ref{part1}), and (\ref{part2}) the matrix ${\bf B}_{\bf 2}^{-1}$ reads
\begin{eqnarray}
{\bf B}_{\bf 2}^{-1} &=& {\bf d}_{\bf 2}^{-1} + (\omega +zt -U)
\underline{u}\underline{u}^{+} \nonumber
\\ &-& \alpha(\underline{u}\underline{u}^{+} -
\underline{u}\underline{w}^{+} - \underline{w}\underline{u}^{+} +
 \underline{w}\underline{w}^{+}) \ .
\end{eqnarray}
The off-diagonal elements of ${\bf B}_{\bf 2}^{-1}$ do not depend explicitly
 on $\underline{k}_1$ and $\underline{k}_2$. In contrast to RES0
 (\ref{invert}), 
however, they are not overall constant but take specific values for 
each block of ${\bf B}_{\bf 2}^{-1}$. To overcome this additional 
complication we introduce the 2$\times$2 matrix
\begin{equation}
{\bf A} = \left[ \begin{array}{cc}
a_1 & a_3 \\ a_3 & a_2 \end{array} \right] = 
\left[ \begin{array}{cc}
\alpha - \omega -zt -U  & -\alpha \\ -\alpha & \alpha \end{array} \right]
\end{equation}
and write ${\bf B}_{\bf 2}^{-1}$ as 
${\bf B}_{\bf 2}^{-1} = {\bf d}_{\bf 2}^{-1} - {\underline{y}}^{+} {\bf A}
\underline{y}$ with $\underline{y}^{+} := \left( \underline{u}, \underline{w}
\right)$. In order to obtain ${\bf B}_{\bf 2}$ we use
an expansion trick  similar to (\ref{funda0}):
\begin{eqnarray} \label{trick}
{\bf B}_{\bf 2} & = & {\bf d}_{\bf 2} ({\bf 1} - \underline{y}^{+}
{\bf A} \underline{y} \, {\bf d}_{\bf 2}) ^{-1} \nonumber \\
 & = & {\bf d}_{\bf 2} ({\bf 1} - \underline{y}^{+} {\bf A} \underline{y} \, 
{\bf d}_{\bf 2} + \underline{y}^{+} {\bf A} \underline{y} \, {\bf d}_{\bf 2}
\underline{y}^{+} {\bf A} \underline{y} \, {\bf d}_{\bf 2} + \ldots)
\nonumber \\
 & = & {\bf d}_{\bf 2} + {\bf d}_{\bf 2} \, \underline{y}^{+} {\bf A}
({\bf 1} - {\bf B}{\bf A})^{-1} \underline{y} \, {\bf d}_{\bf 2} \ ,
\end{eqnarray}
with ${\bf B} := \underline{y} \, {\bf d}_{\bf 2} \, \underline{y}^{+}$
representing the 2$\times$2 matrix
\begin{equation} \label{def:b}
{\bf B} = \left[ \begin{array}{cc} b_1 & b_3 \\ b_3 & b_2 \end{array} \right]
= \left[ \begin{array}{cc} \underline{u}^{+} {\bf d}_{\bf 2} \, 
\underline{u} \ & \underline{u}^{+} {\bf d}_{\bf 2} \, \underline{w} \\
\underline{u}^{+} {\bf d}_{\bf 2} \, \underline{w} \ & 
\underline{w}^{+} {\bf d}_{\bf 2} \, \underline{w} \end{array} \right] \ .
\end{equation}
While inverting the matrix 
\begin{eqnarray} \nonumber
{\bf 1} - {\bf B} {\bf A} &=& \left[ \begin{array}{cc} c_1 & c_3 \\
c_4 & c_2 \end{array} \right]\\
\label{def:c}  &=& \left[ \begin{array}{cc}
1 - a_1 b_1 - a_3 b_3 \ & -a_3 b_1 - a_2 b_3 \ \\
\ -a_1 b_3 - a_3 b_2 & \ 1 - a_2 b_2 - a_3 b_3 \end{array} \right]
\end{eqnarray}
represents a simple algebraic task, the elements of ${\bf B}$ have to be
computed by numerical integration. 
In analogy to $h(\omega) = \underline{v}^{+} {\bf d}_{\bf 1} \, 
\underline{v}$ (see (\ref{hdef})) we introduce 
\begin{eqnarray}
\overline{h} (\omega) &:=& \underline{u}^{+} {\bf d}_{\bf 2} \, \underline{u}
\nonumber \\ \label{hqdef}
& =& |\Lambda|^{-1} \sum_{\underline{k} \in {\rm FS}}
\left[ \delta(\omega -U) + \overline{\gamma} \, \varepsilon(\underline{k})
+ e_1 \right]^{-1}
\end{eqnarray}
with $\overline{\gamma} := \delta - e_1/(zt)$. Just as
$h(\omega)$, $\overline{h} (\omega)$ reduces to an integral over the DOS:
\begin{equation}
\overline{h} (\omega) = 
\int_{\varepsilon_{\rm b}}^{\varepsilon_{\rm F}}
\frac{\rho(\varepsilon) \, d\varepsilon}{\delta (\omega - U) + e_1
+ \overline{\gamma} \varepsilon} = \overline{\gamma}^{\, -1} \,
\overline{G} (\overline{\Omega}) \ ,
\end{equation}
with 
\begin{equation}
\overline{G} (y) := 
\int_{\varepsilon_{\rm b}}^{\varepsilon_{\rm F}}
\frac{\rho (\varepsilon) \, d\varepsilon}{y + \varepsilon} \ , \ 
\overline{\Omega} := \frac{\delta (\omega-U) + e_1}{\overline{\gamma}} \ .
\end{equation}
The symmetry of the DOS with respect to $\varepsilon = 0$
allows to map $\overline{G} (y)$ to the integral $G(y)$ already defined in
(\ref{Gdef}).
Following (\ref{def:b}), the elements of the matrix ${\bf B}$ are 
given by $b_1 = \overline{\gamma}^{\, -1} \overline{G} (\overline{\Omega})$, 
$b_2 = \overline{\gamma}^{\, -1} (zt)^{-2} (e_1 - n \overline{\Omega}
+\overline{\Omega}^2 \, \overline{G} (\overline{\Omega}))$, and
$b_3 = (\overline{\gamma} zt)^{-1} ( n - \overline{\Omega} \,
\overline{G} (\overline{\Omega}))$.

To find the energy of the bound state we have to solve the equation
$(\underline{u}^{+} {\bf B}_{\bf 2} \, \underline{u})^{-1} \doteq 0$.
Starting from (\ref{trick}) and writing $\underline{u}$ formally as
$\underline{u} = \underline{y}^{+} \underline{e}_1$ we obtain
\begin{eqnarray}
\underline{u}^{+} {\bf B}_{\bf 2} \, \underline{u} & = & 
\underline{e}_1^{+} \underline{y} \left( {\bf d}_{\bf 2} + {\bf d}_{\bf 2}
\underline{y} {\bf A} ({\bf 1} - {\bf B}{\bf A})^{-1} \underline{y} \, 
{\bf d}_{\bf 2} \right) \underline{y}^{+} \underline{e}_1 \nonumber \\
 & = & \underline{e}_1^{+} \left( {\bf 1} + {\bf B}{\bf A} ({\bf 1} - 
{\bf B}{\bf A})^{-1} \right) {\bf B} \, \underline{e}_1
\nonumber
\\ &=&   
\underline{e}_1^{+} ({\bf 1} - {\bf B}{\bf A})^{-1} {\bf B} \, \underline{e}_1
\end{eqnarray}
and, using (\ref{def:c}), we finally obtain the equation 
\begin{equation} \label{bound1}
c_4 c_3 - c_1 c_2 \doteq 0
\end{equation}
for the lower edge of the spectrum of $g_{\underline{k}_{\rm b}} (\omega)$.
After inserting all terms (\ref{bound1}) takes the form
\begin{equation} \label{bound2}
p_1 \cdot \overline{G} (\overline{\Omega}) - p_2 \doteq 0 
\end{equation}
with
\begin{eqnarray}
&& p_1 =\nonumber
 \alpha \overline{\gamma} (\overline{\Omega} +zt)^2 - (\omega +zt -U)
\left( \overline{\gamma} (zt)^2 - \alpha(e_1 + n \overline{\Omega})
\right)\\
&& p_2 = \nonumber
 \alpha \overline{\gamma} \left( n (\overline{\Omega} + 2xt
+ e_1) +n^2 (\omega+zt -U) \right) +\overline{\gamma}^2 (zt)^2 \ .
\end{eqnarray}
Making use of the identities $\omega +zt -U = \overline{\gamma}
\delta^{-1} (\overline{\Omega} +zt)$, $e_1 = zt (n - \gamma)$ and
introducing $\chi := \alpha \gamma + zt \overline{\gamma}$,
(\ref{bound2}) simplifies to
\begin{equation} \label{bound3}
\frac{1}{\delta + (\overline{\Omega} +zt) \, \overline{G} (\overline{\Omega})}
\doteq 1 - \frac{zt \chi}{\alpha (\overline{\Omega} +zt)} \ .
\end{equation}
From the definition of $\alpha$ (\ref{def:alpha}) we derive the expression
\begin{equation} 
\label{chi}
\chi = (\Omega+zt) \left( 1 - \frac{1}{n + (\Omega +zt)\, G(\Omega)} \right)
\end{equation}
for $\chi$. We \textit{define} in analogy
\begin{equation}
\label{chiquer}
\overline{\chi} := (\overline{\Omega} +zt) \left( 1 - \frac{1}{\delta +
(\overline{\Omega} +zt)\, \overline{G} (\overline{\Omega})} \right)
\end{equation}
and write (\ref{bound3}) as $\overline{\chi} \doteq zt \chi /\alpha$.
The eliminination of $\alpha$ finally leads to the simple result
\begin{equation} \label{bound4}
\frac{1}{zt} \doteq \frac{\gamma}{\overline{\chi}} + 
\frac{\overline{\gamma}}{\chi} \ .
\end{equation}

The Nagaoka instability line $U_{\rm cr} (\delta)$ is obtained by assuming
$\omega = \varepsilon_{\rm F}$ for a given Fermi energy, calculating $\gamma$, 
$\overline{\gamma}$, and $\chi$ and solve (\ref{bound4}) numerically with
respect to $\overline{\chi}$. Note that $U$ enters  (\ref{bound4}) solely
via $\overline{\Omega}$ and hence via $\overline{\chi}$. To compute
the \textit{optimum spin flip energy} for RES3 for fixed values $U$ 
and $\delta$, we solve (\ref{bound4}) with respect to $\omega$ and 
subtract the Fermi energy $\varepsilon_{\rm F}$ from the solution 
$\omega_0 (U, \delta)$.

\section{General unfrustrated lattice} \label{app:unfrustrated}

In this appendix it will be shown that the formulae derived
in section \ref{sect:resapp} and the formulae (\ref{chi}--\ref{bound4}),
apply to all unfrustrated,
isotropic, homogeneous lattices with nearest neighbor hopping.
In this context, `homogeneous' means that all sites are equivalent;
`isotropic' means that all bonds in all directions are equivalent.
`Unfrustrated' means that the state 
$c_0 := |\Lambda|^{-1/2}\sum_{\underline{i}} a_{\underline{i}}$
is an eigen state of the kinetic Hamiltonian with eigen energy
 $\varepsilon_{\rm b} = - z |t|$ where $t$ is the hopping element
as in (\ref{def:model}) and $z$ is the coordination number.
This requires $t>0$, hence the absence of frustration.
Note that the lattice does not need to be a Bravais lattice. 
The Bethe lattice, however, is not unfrustrated for $z>1$ 
in the above sense
since its lower band edge is $\varepsilon_{\rm b} = - 2 \sqrt{z-1} |t|$
\cite{econo79} and not $\varepsilon_{\rm b} = - z |t|$.

Let us denote by $c^+_\alpha$ the creation operators which
diagonalize the kinetic energy
\begin{equation}
\varepsilon_\alpha c^+_\alpha = [H_{\rm kin},c^+_\alpha]
\end{equation}
and by $a^+_{\underline{j}}$ the site diagonal creation operators.
The unitary transformation between these two bases has
the matrix elements $f_{\alpha,\underline{j}}$
\begin{equation}
c^+_\alpha = \sum\limits_{\underline{j}} 
f_{\alpha,\underline{j}} a^+_{\underline{j}}  \ .
\end{equation}
which implies the expectation values with respect to the
Nagaoka state $|{\cal N}'\rangle$
\begin{mathletters}
\begin{eqnarray}
\label{bed1}
\langle c_{\alpha\uparrow} a^+_{\underline{j}\uparrow}\rangle 
= f^+_{\alpha;\underline{j}}
& 
\quad {\rm for}\quad & \varepsilon_\alpha > \varepsilon_{\rm F}\\
\label{bed2}
\langle a^+_{\underline{j}\uparrow} c_{\beta\uparrow}\rangle  
= f^+_{\beta;\underline{j}}
& 
\quad {\rm for\quad} & \varepsilon_\beta < \varepsilon_{\rm F}
\end{eqnarray}
\end{mathletters}
The homogeneity required implies that
\begin{equation}
\label{bed3}
\sum_\alpha |f_{\alpha,\underline{j}}|^2 \delta(\omega-\varepsilon_\alpha)
= {\rm constant}
\end{equation}
on the lattice, i.e.\ it does not depend on $\underline{j}$.
Unitarity yields furthermore
\begin{equation}
\label{bed4}
\sum_\alpha |f_{\alpha,\underline{j}}|^2 = 1 \ .
\end{equation}

First we address RES0 with the ansatz 
($\varepsilon_\alpha > \varepsilon_{\rm F}$)
\begin{equation}
\label{res0_ansatz_unfrus}
\Phi_\alpha := \sum\limits_{\underline{j}} 
a^{\phantom{+}}_{\underline{j}\uparrow}
c^{+}_{\alpha\uparrow} a^{+}_{\underline{j}\downarrow} |{\cal N}'\rangle 
f^+_{\alpha,\underline{j}} \ .
\end{equation}
The resulting matrix elements are obtained by Wick's theorem and re-expressed
with the help of (\ref{bed1}--\ref{bed4})
\begin{mathletters}
\begin{eqnarray}
\label{xelemu1}
{\bf P}_{\alpha',\alpha} &=&
 n\delta_{\alpha',\alpha} + \sum\limits_{\underline{j}}
|f_{\alpha',\underline{j}}|^2 |f_{\alpha,\underline{j}}|^2  \ ,\\
\label{xelemu2}
{\bf L}_{\uparrow\ \alpha',\alpha} &=& 
(n\varepsilon_\alpha - e_1) \delta_{\alpha',\alpha}\ ,  \\
\label{xelemu3}
{\bf L}_{\downarrow\ \alpha',\alpha} &=& 
-\frac{e_1\varepsilon_\alpha}{zt}\delta_{\alpha',\alpha}
 - t\sum\limits_{\langle \underline{i},\underline{j}\rangle}
 |f_{\alpha',\underline{i}}|^2 |f_{\alpha,\underline{j}}|^2 \ .
\end{eqnarray}
\end{mathletters}
The matrix inversion to be solved is
\begin{mathletters}
\begin{eqnarray}
&&(\omega {\bf P} - {\bf L})^{-1} =
\left( {\bf D}^{-1} + {\bf N} \right)^{-1}\ ,
\\
\label{Ddef}
&&{\bf D}_{\alpha',\alpha} = \delta_{\alpha',\alpha}
(n(\omega-\varepsilon_\alpha) + e_1 + (e_1/zt) \varepsilon_\alpha)^{-1}\ ,
\\
\label{Ndef}
&& {\bf N}_{\alpha',\alpha} = \omega \sum\limits_{\underline{j}}
|f_{\alpha',\underline{j}}|^2 |f_{\alpha,\underline{j}}|^2 
 + t\sum\limits_{\langle \underline{i},\underline{j}\rangle}
 |f_{\alpha',\underline{i}}|^2 |f_{\alpha,\underline{j}}|^2\ . 
\end{eqnarray}
\end{mathletters}
It can be re-expressed with the help of the matrices ${\bf M}$,
 ${\bf A}$, and ${\bf V}$
\begin{mathletters}
\begin{eqnarray}
\label{Mdef}
  {\bf M}_{\underline{i},\underline{j}} & := & \sum\limits_\alpha
  \frac{|f_{\alpha,\underline{i}}|^2 |f_{\alpha,\underline{j}}|^2 }
  {n(\omega-\varepsilon_\alpha) + e_1 + (e_1/zt)
    \varepsilon_\alpha}\ ,\\ 
\label{Adef}
{\bf A}_{\underline{i},\underline{j}} & := &
  \omega\delta_{\underline{i},\underline{j}} + t
  \sum\limits_{\underline{\delta}}
  \delta_{\underline{i}+\underline{\delta},\underline{j}}\ , \\ 
{\bf V}_{\alpha,\underline{j}} & := & |f_{\alpha,\underline{j}}|^2 
\end{eqnarray}
\end{mathletters}
where the $\underline{\delta}$ are all spatial vectors connecting 
nearest neighbors. One obtains
\begin{equation}
\label{unfrus-series}
(\omega {\bf P} - {\bf L})^{-1} = {\bf D} - {\bf D V A}
\left( \sum_{n=0}^\infty (-{\bf MA})^n \right){\bf V}^+{\bf D} \ .
\end{equation}
The key observation at this stage is that the vector $\underline{u}$
with $u_{\underline{j}}=|\Lambda|^{-1/2}$ is an eigenvector both of 
the matrices ${\bf M}$ and ${\bf A}$. The corresponding eigenvalue
for ${\bf M}$ is found with the help of (\ref{bed3})
\begin{equation}
\label{hdef2}
h(\omega) = \frac{1}{L}\sum\limits_{\alpha,\underline{i},\underline{j}}
\frac{|f_{\alpha,\underline{i}}|^2 |f_{\alpha,\underline{j}}|^2 }
  {n(\omega-\varepsilon_\alpha) + e_1 + (e_1/zt)
    \varepsilon_\alpha}
\end{equation}
which simplifies due to (\ref{bed4}) in the end to the form
(\ref{hdef}). The corresponding eigenvalue of ${\bf A}$ is
$\omega+zt = \omega- \varepsilon_{\rm b}$. So the series in 
(\ref{unfrus-series})
yields a vanishing denominator for
$0=1+(\omega-\varepsilon_{\rm b})h(\omega)$. Thus we derived (\ref{res0}) for
a much broader class of lattices.

The equations for RES1 and RES2 follow in analogy to the derivation
in sect.\ \ref{sect:resapp}B. 
The ansatz RES1 is identical to (\ref{varib}) for 
$\underline{k}_{\rm b}=\underline{0}$ and the additional matrix
elements are the same as in (\ref{xelem2}) once 
$\varepsilon_{\underline{k}}$ is replaced by $\varepsilon_\alpha$.
An important point to note is that the homogeneity (\ref{bed3})
ensures that $N_\alpha$ couples indeed to the constant eigenvector
$\underline{u}$
\begin{equation}
({\bf V}^+{\bf DN})_{\underline{j}} =
(\delta-h(\omega)n(\omega-\varepsilon_{\rm b}))u_{\underline{j}}
\end{equation}
for which the series summation in (\ref{unfrus-series}) was achieved.

For the ansatz RES2 we work with (\ref{varic}) for 
$\underline{k}_{\rm b}=\underline{0}$ and find the matrix elements
(\ref{xelem3}) after replacing $\varepsilon_{\underline{k}}$
by $\varepsilon_\alpha$. Using
\begin{equation}
({\bf V}^+{\bf DN})_{\underline{j}} =
 y u_{\underline{j}}
\end{equation}
with $y$ as in (\ref{ydef}), we obtain again (\ref{res2}) as condition
for the variational spin flip energy.

Let us now turn to RES3. We use ($\varepsilon_\beta < \varepsilon_{\rm F}$)
\begin{equation}
\label{res3_unfrus}
\Psi_\beta = \sum_{\underline{j}}
a^{{+}}_{\underline{j}\uparrow}
c^{\phantom{+}}_{\beta\uparrow} a^{+}_{\underline{j}\downarrow} 
|{\cal N}'\rangle 
f_{\beta,\underline{j}} 
\end{equation}
in analogy to (\ref{varid}) for the doubly occupied states.
The matrices ${\bf D_1}$ for $\Phi_\alpha$
and ${\bf D_2}$ for $\Psi_\beta$ as in (\ref{block2}) are given by
\begin{mathletters}
\begin{eqnarray}\nonumber
&&({\bf D_1})_{\alpha',\alpha} =
\delta_{\alpha',\alpha} (n(\omega-\varepsilon_\alpha)
+e_1 + e_1\varepsilon_\alpha/(zt))
+\\ \label{D1def_gen}
&&\quad \omega \sum\limits_{\underline{j}}
|f_{\alpha',\underline{j}}|^2 |f_{\alpha,\underline{j}}|^2 
+ t \sum\limits_{\langle\underline{j},\underline{j}\rangle}
|f_{\alpha',\underline{i}}|^2 |f_{\alpha,\underline{j}}|^2\ ,
\end{eqnarray}
\begin{eqnarray}\nonumber
&&({\bf D_2})_{\beta',\beta} =
\delta_{\beta',\beta} (\delta(\omega-U+\varepsilon_\beta)
+e_1 - e_1\varepsilon_\beta/(zt))
+\\ \label{D2def_gen}
&&\quad \omega \sum\limits_{\underline{j}}
|f_{\beta',\underline{j}}|^2 |f_{\beta,\underline{j}}|^2 
+ t \sum\limits_{\langle\underline{j},\underline{j}\rangle}
|f_{\beta',\underline{i}}|^2 |f_{\beta,\underline{j}}|^2 
\end{eqnarray}
\end{mathletters}
where ${\bf D_1}$ can be read off from (\ref{Ddef},\ref{Ndef}) and
${\bf D_2}$ is analogous for the states $\Psi_\beta$.

The matrix ${\bf N}$, which couples the doubly and the non-doubly
occupied subspaces (see (\ref{block2})), is obtained again 
via Wick's theorem and with (\ref{bed2},\ref{bed3})
\begin{eqnarray}\nonumber
{\bf N}_{\alpha,\beta} &=&
t \sum\limits_{\langle \underline{i},\underline{j}\rangle}
\left(
f^+_{\alpha,\underline{j}} f^{\phantom{+}}_{\alpha,\underline{i}} 
|f_{\beta,\underline{i}}|^2
-|f_{\alpha,\underline{i}}|^2 
f^+_{\beta,\underline{j}} f^{\phantom{+}}_{\beta,\underline{i}} 
\right)
\\ \label{Ndef_gen}
&& + t \sum\limits_{\langle\underline{i},\underline{j}\rangle}
\left(
f^+_{\alpha,\underline{i}} f^{\phantom{+}}_{\alpha,\underline{j}}
f^+_{\beta,\underline{j}} f^{\phantom{+}}_{\beta,\underline{i}}
-|f_{\alpha,\underline{j}}|^2
|f_{\beta,\underline{i}}|^2
\right)\ .
\end{eqnarray}
In order to re-express the inverse matrix ${\bf B _2}=
({\bf D_2}- {\bf N}^+ {\bf D_1}^{-1} {\bf N})^{-1}$
we define
\begin{mathletters}
\begin{equation}
({\bf C_{1}})_{\underline{i},\underline{\delta}';
\underline{j},\underline{\delta}}
:= \sum\limits_{\alpha',\alpha} 
f^+_{\alpha',\underline{i}+\underline{\delta}'}
f^{\phantom{+}}_{\alpha',\underline{i}}
({\bf D_1})_{\alpha',\alpha}
f^+_{\alpha',\underline{j}+\underline{\delta}j}
f^{\phantom{+}}_{\alpha,\underline{j}}\ .
\end{equation}
\begin{equation}
({\bf C_{2}})_{\underline{i},\underline{\delta}';
\underline{j},\underline{\delta}}
:= \sum\limits_{\beta',\beta} 
f^+_{\beta',\underline{i}+\underline{\delta}'}
f^{\phantom{+}}_{\beta',\underline{i}}
({\bf D_2})_{\beta',\beta}
f^+_{\beta',\underline{j}+\underline{\delta}j}
f^{\phantom{+}}_{\beta,\underline{j}}\ ,
\end{equation}
\begin{eqnarray}\nonumber
{\bf E}_{\underline{i},\underline{\delta}';
\underline{j},\underline{\delta}}
&:=& 
-t \delta_{\underline{i},\underline{j}} 
(\delta_{\underline{\delta}',\underline{0}} -
\delta_{\underline{\delta},\underline{0}})
+t \delta_{\underline{i}-\underline{\delta}',\underline{j}}
\delta_{\underline{\delta},-\underline{\delta}'}
(1-\delta_{\underline{\delta},\underline{0}})
\\ \label{Edef}
&-&t\sum\limits_{\underline{\delta}''}
\delta_{\underline{i}+\underline{\delta}'',\underline{j}}
\delta_{\underline{\delta}',\underline{0}}
\delta_{\underline{\delta},\underline{0}}
\ ,
\end{eqnarray}
\begin{equation}
{\bf V}_{\underline{j},\delta;\beta} := 
f^+_{\beta,\underline{j}+\underline{\delta}}
f^{\phantom{+}}_{\beta,\underline{j}}
\end{equation}
\end{mathletters}
where the spatial vectors $\underline{\delta}$, $\underline{\delta}'$,
and  $\underline{\delta}''$
link nearest neighbors or equal $\underline{0}$.
The result is
\begin{eqnarray} \nonumber
&&{\bf B_2} = {\bf D_2}^{-1} \\
\label{geomseries}&+&
 \sum\limits_{n=0}^\infty
{\bf D_2}^{-1}
{\bf V}^+{\bf E}^+{\bf C_1}{\bf E}
({\bf C_2}{\bf E}^+{\bf C_1}{\bf E})^n{\bf VD_2}^{-1}
\end{eqnarray}
where we once again focus on the geometric series.
In slight extension of the situation for RES0-2 we do not
guess one common eigenvector of ${\bf E}$, ${\bf C_1}$, and
${\bf C_2}$ but a two-dimensional subspace spanned by 
$\underline{u}$ and $\underline{v}$.
The vectors
 are defined by  $u_{\underline{i},\underline{0}}:= |\Lambda|^{-1/2}$
and zero otherwise, 
and by $v_{\underline{i},\underline{\delta}\neq\underline{0}}:= 
(z|\Lambda|)^{-1/2}$ and zero otherwise.

Straightforward calculation shows
\begin{mathletters}
\label{Eform}
\begin{eqnarray}
{\bf E} \underline{u} &=& -zt \underline{u} + \sqrt{z}t\underline{v}\ ,\\
{\bf E} \underline{v} &=& -\sqrt{z}t \underline{u} + t\underline{v}
\end{eqnarray}
\end{mathletters}
which can be summarized in 
\begin{equation}
\label{esimple}
{\bf E} = -zt\ \underline{a}\ \underline{b}^+ 
\end{equation}
with $\underline{a}:= \underline{u}-\underline{v}/\sqrt{z}$
and $\underline{b}:= \underline{u}+\underline{v}/\sqrt{z}$.
The matrix elements of ${\bf C_1}$ with respect to
$\underline{u}$ and $\underline{v}$ in obvious notation are
\begin{mathletters}
\label{C1}
\begin{eqnarray}\nonumber
{\bf C_1}^{uu} &=& \frac{1}{|\Lambda|}\sum\limits_{\alpha',\alpha}
({\bf D_1})_{\alpha',\alpha}
\\
&=& \frac{h_0}{1+(\omega-\varepsilon_{\rm b})h_0}\ ,
\end{eqnarray}
\begin{eqnarray}\nonumber
{\bf C_1}^{vv} &=& \frac{1}{z t^2 |\Lambda|}\sum\limits_{\alpha',\alpha}
\varepsilon_{\alpha'}({\bf D_1})_{\alpha',\alpha}\varepsilon_{\alpha}
\\
&=& \frac{1}{z t^2}
\left(h_2 - \frac{h_1^2(\omega-\varepsilon_{\rm b})}{1+(
\omega-\varepsilon_{\rm b})h_0}\right)\ ,
\end{eqnarray}
\begin{eqnarray}\nonumber
{\bf C_1}^{uv} =  {\bf C_1}^{vu} &=&
 \frac{-1}{\sqrt{z}t |\Lambda|}\sum\limits_{\alpha',\alpha}
\varepsilon_{\alpha'}({\bf D_1})_{\alpha',\alpha}
\\
&=& \frac{-1}{\sqrt{z}t} \frac{h_1}{1+(\omega-\varepsilon_{\rm b})h_0} \ ,
\end{eqnarray}
\end{mathletters}
where we use the generalization of (\ref{hdef}) ($h_0=h$)
\begin{equation}
\label{hdef-extend}
h_n := \int\limits_{\varepsilon_{\rm F}}^{\varepsilon_{\rm t}}
\frac{\varepsilon^n \rho(\varepsilon)d\varepsilon}{\Omega-\gamma\varepsilon}
\ .
\end{equation}
It is useful to keep the following relations in mind
\begin{mathletters}
\begin{eqnarray}
h_1 &=& (-\delta + (n\omega+e_1)h_0)/\gamma\ , \\
h_2 &=& (e_1 + (n\omega+e_1)h_1)/\gamma \ .
\end{eqnarray}
\end{mathletters}
For ${\bf C_2}$ very similar equations are derived after replacing
$\alpha$ by $\beta$, i.e.\ by changing the summation over the unoccupied levels
to a summation over the occupied levels
\begin{mathletters}
\label{C2}
\begin{eqnarray}
{\bf C_2}^{uu} &=& \frac{\bar h_0}{1+(\omega-U-\varepsilon_{\rm b})\bar h_0}\ ,
\\
{\bf C_2}^{vv} &=& \frac{1}{z t^2}
\left(\bar h_2 - \frac{\bar h_1^2(\omega-U-\varepsilon_{\rm b})}
{1+(\omega-U-\varepsilon_{\rm b})\bar h_0}\right)\ ,
\\
{\bf C_2}^{uv} &=&  {\bf C_2}^{vu} =
 \frac{-1}{\sqrt{z}t} \frac{\bar h_1}{1+(\omega-U-
\varepsilon_{\rm b})\bar h_0} \ ,
\end{eqnarray}
\end{mathletters}
where the generalization of (\ref{hqdef}) ($\bar h_0 = \bar h$)
\begin{equation}
\bar h_n := \int\limits^{\varepsilon_{\rm F}}_{\varepsilon_{\rm b}}
\frac{\varepsilon^n \rho(\varepsilon)d\varepsilon}{\bar\Omega+
\bar\gamma\varepsilon}
\end{equation}
is used. The following relations hold
\begin{mathletters}
\begin{eqnarray}
\bar h_1 &=& (n - (\delta(\omega-U)+e_1)\bar h_0)/\bar\gamma\ , \\
\bar h_2 &=& (e_1 - (\delta(\omega-U)+e_1)\bar h_1)/\bar\gamma \ .
\end{eqnarray}
\end{mathletters}

Due to the particularly simple form of ${\bf E}$ in (\ref{esimple})
all we need to do is to calculate
\begin{mathletters}
\begin{eqnarray}
c_1&:=& \underline{a}^+ {\bf C_1} \underline{a}
= \frac{zt}{1+(\omega-\varepsilon_{\rm b})h_0}\Big(h_0+\frac{2h_1}{zt}
\nonumber \\
&&+\frac{1}{zt}(h_2+(\omega-\varepsilon_{\rm b})(h_2h_0-h_1^2))  \Big)\ ,\\
c_2&:=& \underline{b}^+ {\bf C_2} \underline{b}
= \frac{zt}{1+(\omega-U-\varepsilon_{\rm b})\bar h_0}
\Big(\bar h_0-\frac{2\bar h_1}{zt}
\nonumber \\
&&+\frac{1}{zt}(\bar h_2+(\omega-U-\varepsilon_{\rm b})
(\bar h_2\bar h_0-\bar h_1^2))\Big)\ .
\end{eqnarray}
\end{mathletters}
The condition for the singularity of (\ref{geomseries}) reads now
\begin{equation}
1\doteq c_1 c_2
\end{equation}
which is equivalent to (\ref{bound4}) as can be shown by some 
tedious, but straightforward calculation. 
Thus we have completed the proof that the equations for RES0--3
derived in the main text for hypercubic lattices hold for all unfrustrated,
isotropic, homogeneous lattices with nearest neighbor hopping. 
Only the coordination number and the DOS enter the evaluation of
the RES ansatzes.

\section{Triangular lattice} \label{app:tri}
For the triangular lattice with $t < 0$ the lower band edge
$\varepsilon_{\rm b} = -3|t|$ is reached at $\underline{k}_{\rm b} =
(4\pi/3, 0)$. Since $\underline{k}_{\rm b} \neq \underline{0}$, the 
integral
\begin{equation}
h(\omega) = \left\langle \frac{1}{n[\omega - \varepsilon (\underline{k})]
+ e_1 [1 - \varepsilon (\underline{k} - \underline{k}_{\rm b})/(zt)]}
\right\rangle_{\underline{k} \in {\rm BZ} \setminus {\rm FS}}
\end{equation}
cannot be mapped onto a one-dimensional integral over the DOS
but has to be evaluated explicitly in momentum space. 

The optimum spin flip energy for RES0 for a given hole density $\delta$
follows from the solution $\omega_{0}$ of the equation 
$1 + (\omega - \varepsilon_{\rm b}) h(\omega) \doteq 0$ as
$\Delta e_{\infty} (\delta) = \omega_{0} - \varepsilon_{\rm F}$
(see sect.\ \ref{sect:resapp}). To obtain the Fermi energy corresponding to
the critical hole density $\delta_{\rm cr}$ the equation
$1 + (\varepsilon_{\rm F} - \varepsilon_{\rm b}) h(\varepsilon_{\rm F})
\doteq 0$ has to be solved numerically.

For RES1, (\ref{res1}) holds also for 
$\underline{k}_{\rm b} \neq \underline{0}$, since 
$|\Lambda|^{-1} \sum_{\underline{k}} \varepsilon (\underline{k} -
\underline{k}_{\rm b}) = \varepsilon_{\rm b} \, e_{1}/(zt)$ due to the
symmetry of the lattice. Calculating $\underline{N}^{+} {\bf D}_{\bf 1}^{-1} 
\underline{N}$ for RES2, however, the integrals
\begin{equation} \label{def:h1h2}
h_{n} = \left\langle \frac{\varepsilon^{n} (\underline{k})}{n[\omega 
- \varepsilon (\underline{k})] + e_1 [1 - \varepsilon (\underline{k} 
- \underline{k}_{\rm b})/(zt)]} \right\rangle_{\underline{k} \in {\rm BZ} 
\setminus {\rm FS}}
\end{equation}
which for $\underline{k}_{\rm b}=\underline{0}$ simplify to (\ref{hdef-extend})
 have to be computed for $n = 1,2$. Although the
outline of the derivation remains unchanged, this causes some differences in
the analytic expressions for the optimum spin flip energy and the Nagaoka 
instability line compared with the case $\underline{k}_{\rm b} 
= \underline{0}$ (see sect.\ \ref{sect:resapp}).

Evaluating the full resolvent ansatz RES3, the product in the 
second line of (\ref{factor1}) can be written as
\begin{eqnarray}
&&\varepsilon_{\rm b}^{2} \left( 1 + \frac{\varepsilon(\underline{k}_1)}
{\varepsilon_{\rm b}} - \frac{\varepsilon (\underline{q}_1)}{\varepsilon_{\rm 
b}} \left( 1 + \frac{\varepsilon (\underline{k}_1)}{4 \varepsilon_{\rm b}}
\right) \right) \times
\nonumber \\
&&\qquad \left( 1 + \frac{\varepsilon(\underline{k}_2)}
{\varepsilon_{\rm b}} - \frac{\varepsilon (\underline{q}_2)}{\varepsilon_{\rm 
b}} \left( 1 + \frac{\varepsilon (\underline{k}_2)}{4 \varepsilon_{\rm b}}
\right) \right) \ , 
\end{eqnarray}
making use of $\varepsilon_{\rm b} = -z|t|/2$. The permutation symmetry with 
respect to the primitive lattice vectors which is essential for the 
factorization $\varepsilon(\underline{k} - \underline{q}) = 
-\varepsilon(\underline{k}) \varepsilon(\underline{q})/(zt)$ holds also for
the triangular lattice. The matrix ${\bf N}^{+} {\bf D}_{\bf 1}^{-1}
{\bf N}$ is calculated to be
\begin{equation}
{\bf N}^{+} {\bf D}_{\bf 1}^{-1} {\bf N} = \alpha_1 \underline{u}
\underline{u}^{+} + \alpha_2 (\underline{u} \underline{w}^{+} + \underline{w}
\underline{u}^{+}) + \alpha_2 \underline{w} \underline{w}^{+}
\end{equation}
with
\begin{mathletters}
\begin{eqnarray} 
&&\alpha_1 = \varepsilon_{\rm b}^{2} - 2\varepsilon_{\rm b} h_{1}  + 
h_2  \label{def:alpha1} = H [\varepsilon_{\rm b} h  - h_1  ]^2,
\\ \nonumber
&&\alpha_2 = \varepsilon_{b}^2 h  - \frac{5}{4} \varepsilon_{\rm b}
h_1   + \frac{1}{4} h_2   
\\ \label{def:alpha2}
&&- H\left[\varepsilon_{\rm b}^2
h^2   - \frac{5}{4} \varepsilon_{\rm b} h  h_1  
+ \frac{1}{4} h_1^2  \right]
\\ \nonumber
&&\alpha_3 = \varepsilon_{b}^2 h  = \frac{1}{2} \varepsilon_{\rm b}
h_1   + \frac{1}{16} h_2    \\
 \label{def:alpha3}
&&- H
\left[\varepsilon_{\rm b}^2 h^2   - \frac{1}{2} \varepsilon_{\rm b} 
h  h_1   + \frac{1}{16} h_1^2  \right].
\end{eqnarray}
\end{mathletters}
In (\ref{def:alpha1}) -- (\ref{def:alpha3}), $H$ is a short-hand
notation for $(\omega - \varepsilon_{\rm b})/[1 + 
(\omega-\varepsilon_{\rm b}) h(\omega)]$. The method developed in appendix
\ref{app:res3} to calculate $\underline{u}^{+} {\bf B}_{\bf 2}
\underline{u}$ is applicable also for the triangular lattice up to
eq.\ (\ref{bound1}) which yields the optimum spin flip energy for RES3.

The elements of the 2$\times$2 matrices ${\bf A}$ and ${\bf B}$ are given by
$a_1 = \alpha_1 - (\omega - \varepsilon_{\rm b} -U)$, 
$a_2 = \alpha_3$, $a_3 = \alpha_2$, 
$b_1 = \overline{h} (\omega)$, $b_2 = \overline{h}_2 (\omega)/
\varepsilon_{\rm b}^2$, $b_3 = \overline{h}_1 (\omega)/\varepsilon_{\rm b}$
with $\overline{h}_1 (\omega)$ and $\overline{h}_2 (\omega)$ defined in
analogy to (\ref{def:h1h2}) as integrals over the Fermi sphere.

\section{Kagome lattice} \label{app:kagome}

To prepare  the derivation of the ansatzes RES0 -- RES3 for the frustrated
Kagome lattice we diagonalize the one-particle problem explicitly.
\begin{figure}[htb]
\begin{center}\includegraphics[width=8cm]{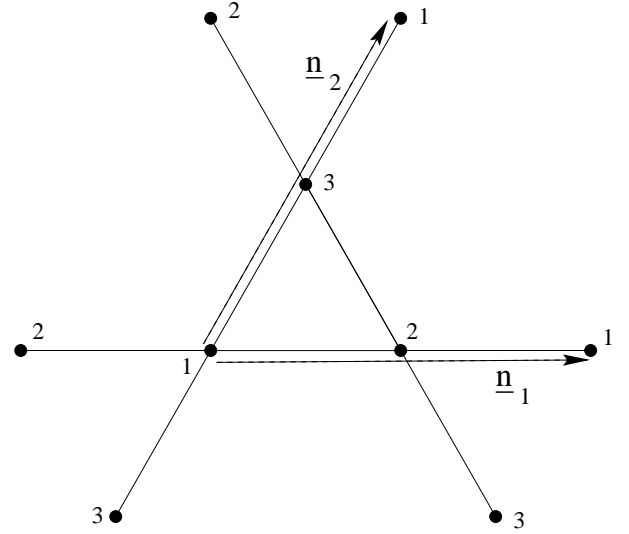}\end{center}
 \caption{ \label{fig:kagome} 
   Segment of the kagome lattice. The vectors are used in the main text.
   The numbers refer to the three sites in each unit cell of this non-Bravais
   lattice.}
\end{figure} \noindent
Since we deal with a non-Bravais lattice with three sites per unit cell
we have to solve a 3$\times$3 eigenvalue problem with
\begin{equation}
\label{kagphas}
f_{\alpha,{\underline j}} = \exp(i {\underline k}{\underline j}) \
\phi_{\alpha,\tau({\underline j})} 
\end{equation}
where $\tau({\underline j})\in\{1,2,3\}$ denotes the sub-lattice to which site
 $\underline j$
belongs.
 The one-particle Hamiltonian acting on $\phi_{\alpha,\tau}$ becomes
\begin{equation}
\label{hammatrix}
h({\underline k}) = 2t\left(  
\begin{array}{ccc}
0 & \cos({\underline k}{\underline n}_1/2) & 
\cos(\frac{{\underline k}{\underline n}_2}{2}) \\
\cos(\frac{{\underline k}{\underline n}_1}{2}) & 0 & 
\cos(\frac{{\underline k}({\underline n}_2 - {\underline n}_1)}{2}) \\
\cos(\frac{{\underline k}{\underline n}_2}{2}) & 
\cos(\frac{{\underline k}({\underline n}_2 - {\underline n}_1)}{2}) & 0
\end{array}   \right)
\end{equation}
where we used the unit vectors ${\underline n}_{1}$ and  
${\underline n}_{2}$ as shown
in fig.\ (\ref{fig:kagome}).
The secular equation of (\ref{hammatrix}) is
\begin{equation}
\label{seckag}
0 = (-2t+\lambda)(\lambda^2+2t\lambda -2t^2+
t\varepsilon_\triangle({\underline k}) )\ ,
\end{equation}
where $\varepsilon_\triangle({\underline k})$ is the triangular dispersion
(\ref{dreieck_dispers}). From the secular equation one deduces
(\ref{kagdispers}) easily. More important for the following is the
observation that $h({\underline k})$ can be diagonalized by an orthogonal,
 i.e.\
real, transformation since it is real symmetric. Thus the phase of 
$f_{\alpha,{\underline j}}$ is completely given by the plane wave factor
$ \exp(i {\underline k}{\underline j})$ in (\ref{kagphas}).

Since we wish to treat the frustrated case ($t<0$) we modify the ansatz
(\ref{res0_ansatz_unfrus}) by introducing an additional phase factor
$\lambda_{\underline{j}}$ depending only on the sub-lattice and being unity
on sub-lattice 1, $\exp(2\pi i/3)$ on sub-lattice 2, and $\exp(-2\pi i/3)$
 on sub-lattice 3
\begin{equation}
\label{res0_ansatz_kag}
\Phi_\alpha := \sum\limits_{\underline{j}} 
a^{\phantom{+}}_{\underline{j}\uparrow}
c^{+}_{\alpha\uparrow} a^{+}_{\underline{j}\downarrow} |{\cal N}'\rangle 
f^+_{\alpha,\underline{j}} \lambda^+_{\underline{j}} \ .
\end{equation}
The resulting matrix elements ${\bf P}_{\alpha',\alpha}$ and 
${\bf L}_{\uparrow\ \alpha',\alpha}$ are the same as in 
(\ref{xelemu1},\ref{xelemu2})
since the phase factor cancels at each site. But 
${\bf L}_{\downarrow\ \alpha',\alpha}$ does change into
\begin{equation}
\label{lrunter}
{\bf L}_{\downarrow\ \alpha',\alpha} = 
\frac{e_1\varepsilon_\alpha}{2zt}\delta_{\alpha',\alpha}
 - t\sum\limits_{\langle \underline{i},\underline{j}\rangle}
 |f_{\alpha',\underline{i}}|^2 |f_{\alpha,\underline{j}}|^2 
\lambda^+_{\underline{i}} \lambda^{\phantom +}_{\underline{j}} \ .
\end{equation}
The change in the second term is obvious. The change in the first
term $A_1$ is less trivial. In a first step one obtains
\begin{equation}
\label{a1}
A_1 = -\frac{e_1}{zt}\delta_{\alpha',\alpha} 
\sum\limits_{\underline{k},\underline{\delta}} 
f^+_{\alpha,{\underline j}} f^{\phantom +}_{\alpha,{\underline j}+
\underline{\delta}}
\lambda^+_{\underline{j}} \lambda^{\phantom +}_{\underline{j}+
\underline{\delta}}\ .
\end{equation}
Transforming the terms of the sum like ${\underline \delta} \rightarrow
-{\underline \delta}$ and ${\underline j} \rightarrow
{\underline j} + {\underline \delta}$ leads to 
\begin{eqnarray}
f^+_{\alpha,{\underline j}} f^{\phantom +}_{\alpha,{\underline j}+
\underline{\delta}}
\lambda^+_{\underline{j}} \lambda^{\phantom +}_{\underline{j}+
\underline{\delta}}
&\to& \nonumber
f^+_{\alpha,{\underline j}+ \underline{\delta}} 
f^{\phantom +}_{\alpha,{\underline j}+ 2\underline{\delta}}
\lambda^{\phantom +}_{\underline{j}} \lambda^+_{\underline{j}+
\underline{\delta}} \\
&=& f^+_{\alpha,{\underline j}} f^{\phantom +}_{\alpha,{\underline j}+
\underline{\delta}}
\lambda^{\phantom +}_{\underline{j}} \lambda^+_{\underline{j}+
\underline{\delta}}\ .
\end{eqnarray}
The last equality holds since $\phi_{\alpha,\tau({\underline j})}$ in
 (\ref{kagphas})
 is real. Hence only the real part of 
$\lambda^+_{\underline{j}} \lambda^{\phantom +}_{\underline{j}+
\underline{\delta}}$
in (\ref{a1}) matters. It is -1/2 leading thus to the first term in
(\ref{lrunter}).

From the matrix elements (\ref{xelemu1},\ref{xelemu2},\ref{lrunter}) we
find the relations which are analogous to 
(\ref{Ddef},\ref{Ndef},\ref{Mdef},\ref{Adef})
\begin{mathletters}
\begin{eqnarray}
{\bf D}_{\alpha',\alpha} &=& \delta_{\alpha',\alpha}
(n(\omega-\varepsilon_\alpha) + e_1 - (e_1/2zt) \varepsilon_\alpha)^{-1}\ ,
\\
{\bf N}_{\alpha',\alpha} &=& \omega \sum\limits_{\underline{j}}
|f_{\alpha',\underline{j}}|^2 |f_{\alpha,\underline{j}}|^2 
 \nonumber \\
\label{Ndef_kag}
&+& t\sum\limits_{\langle \underline{i},\underline{j}\rangle}
 |f_{\alpha',\underline{i}}|^2 |f_{\alpha,\underline{j}}|^2
\lambda^+_{\underline{i}} \lambda^{\phantom +}_{\underline{j}}\ , \\
  {\bf M}_{\underline{i},\underline{j}} & := & \sum\limits_\alpha
  \frac{|f_{\alpha,\underline{i}}|^2 |f_{\alpha,\underline{j}}|^2 }
  {n(\omega-\varepsilon_\alpha) + e_1 - (e_1/2zt)
    \varepsilon_\alpha}\ ,\\ 
{\bf A}_{\underline{i},\underline{j}} & := &
  \omega\delta_{\underline{i},\underline{j}} + t
  \sum\limits_{\underline{\delta}}
  \delta_{\underline{i}+\underline{\delta},\underline{j}} 
\lambda^+_{\underline{i}} \lambda^{\phantom +}_{\underline{j}}\ .
\end{eqnarray}
\end{mathletters}
The vector $\underline u$ is again an eigenvector of the matrices ${\bf M}$
and ${\bf A}$. Its eigenvalue for ${\bf M}$ is in analogy to 
(\ref{hdef2}) identical to (\ref{hdef}) with the adapted definition
\begin{equation}
\label{gamkag}
\gamma_{\rm K} = n + e_1/(2zt)\ .
\end{equation}
The eigenvalue for ${\bf A}$ is $\omega -zt/2= 
\omega- \varepsilon_{\rm b}$ as before.
So the series in  (\ref{unfrus-series})
yields a vanishing denominator for
$0=1+(\omega-\varepsilon_{\rm b})h(\omega)$ with $h(\omega)$ as
in (\ref{hdef}) with $\gamma$ (\ref{gamdef}) replaced by $\gamma_{\rm K}$
(\ref{gamkag}). So the DOS, the lower band edge $\varepsilon_{\rm b}$,
and $\gamma_{\rm K}$ are the only quantities to be changed in order that RES0
 (\ref{res0} applies to the frustrated kagome lattice.

For RES1 the ansatz reads 
\begin{equation}
| \Psi_1 \rangle
 :=  |\Lambda|^{-1/2}\sum\limits_{\underline{i}}
 \exp(i\underline{k}_{\rm b} \underline{i})
a_{\underline{i}\uparrow}^{+}
a_{\underline{i}\uparrow}^{\phantom{+}}
a_{\underline{i}\downarrow}^{+} | {\cal N}' \rangle  
\lambda^+_{\underline{i}}
\end{equation}
in extension of (\ref{varib}). The 
resulting condition is identical to (\ref{res1}) with the adapted
$\gamma_{\rm K}$ in (\ref{gamkag}) and, of course, 
$\varepsilon_{\rm b} = zt/2$.

For RES2 the ansatz reads
\begin{equation}
| \Psi_2 \rangle
 :=  |\Lambda|^{-1/2}\sum\limits_{<\underline{i}\underline{j}>}
 \exp(i\underline{k}_{\rm b} \underline{i})
a_{\underline{i}\uparrow}^{+}
a_{\underline{j}\uparrow}^{\phantom{+}}
a_{\underline{i}\downarrow}^{+} | {\cal N}' \rangle \lambda^+_{\underline{i}} 
\end{equation}
yielding again condition (\ref{res2}) with the adapted quantities, in
particular $\gamma_{\rm K}':= e_1 + e_2/(2zt)$.

The ansatz for the doubly occupied states in RES3 is the extension
of (\ref{res3_unfrus})
\begin{equation}
\Psi_\beta = \sum_{\underline{j}}
a^{{+}}_{\underline{j}\uparrow}
c^{\phantom{+}}_{\beta\uparrow} a^{+}_{\underline{j}\downarrow} 
|{\cal N}'\rangle 
f_{\beta,\underline{j}}  \lambda^+_{\underline{j}} \ .
\end{equation}
The relations analogous to 
(\ref{D1def_gen},\ref{D2def_gen},\ref{Ndef_gen},\ref{Edef}) read
\begin{mathletters}
\begin{eqnarray}\nonumber
&&({\bf D_1})_{\alpha',\alpha} = 
\delta_{\alpha',\alpha} (n(\omega-\varepsilon_\alpha)
+e_1 - e_1\varepsilon_\alpha/(2zt))
+\\
&&\  \omega \sum\limits_{\underline{j}}
|f_{\alpha',\underline{j}}|^2 |f_{\alpha,\underline{j}}|^2 
+ t \sum\limits_{\langle\underline{i},\underline{j}\rangle}
|f_{\alpha',\underline{i}}|^2 |f_{\alpha,\underline{j}}|^2
 \lambda^+_{\underline{i}} \lambda^{\phantom +}_{\underline{j}} 
,\\ \nonumber
&&({\bf D_2})_{\beta',\beta} =
\delta_{\beta',\beta} (\delta(\omega-U+\varepsilon_\beta)
+e_1 + e_1\varepsilon_\beta/(2zt))
+\\
&&\ \omega \sum\limits_{\underline{j}}
|f_{\beta',\underline{j}}|^2 |f_{\beta,\underline{j}}|^2 
+ t \sum\limits_{\langle\underline{i},\underline{j}\rangle}
|f_{\beta',\underline{i}}|^2 |f_{\beta,\underline{j}}|^2 
 \lambda^+_{\underline{i}} \lambda^{\phantom +}_{\underline{j}} 
,\\ \nonumber
&&{\bf N}_{\alpha,\beta} =
t \sum\limits_{\underline{i},\underline{j}}
\left(
f^+_{\alpha,\underline{j}} f^{\phantom{+}}_{\alpha,\underline{i}} 
|f_{\beta,\underline{i}}|^2
-|f_{\alpha,\underline{i}}|^2 
f^+_{\beta,\underline{j}} f^{\phantom{+}}_{\beta,\underline{i}} 
\right) +
\\
&&\ t \sum\limits_{\underline{i},\underline{j}}
\big(\frac{-1}{2}
f^+_{\alpha,\underline{i}} f^{\phantom{+}}_{\alpha,\underline{j}}
f^+_{\beta,\underline{j}} f^{\phantom{+}}_{\beta,\underline{i}}
-|f_{\alpha,\underline{j}}|^2 |f_{\beta,\underline{i}}|^2 
\lambda^+_{\underline{i}} \lambda^{\phantom +}_{\underline{j}} 
\big) ,\\ \nonumber
&&{\bf E}_{\underline{i},\underline{\delta}';
\underline{j},\underline{\delta}} = 
-t \delta_{\underline{i},\underline{j}} 
(\delta_{\underline{\delta}',\underline{0}} -
\delta_{\underline{\delta},\underline{0}})
- \frac{t}{2} \delta_{\underline{i}-\underline{\delta}',\underline{j}}
\delta_{\underline{\delta},-\underline{\delta}'}
(1-\delta_{\underline{\delta},\underline{0}})
\\ \label{Edef_kag}
&&\  - t\sum\limits_{\underline{\delta}''}
\delta_{\underline{i}+\underline{\delta}'',\underline{j}}
\delta_{\underline{\delta}',\underline{0}}
\delta_{\underline{\delta},\underline{0}}
\lambda^+_{\underline{i}} \lambda^{\phantom +}_{\underline{j}} \ .
\end{eqnarray}
\end{mathletters}

So far the analogy to the treatment of unfrustrated lattices is
perfect once the different form of $\varepsilon_{\rm b}$, $\gamma_{\rm K}$, and
of $\bar\gamma_{\rm K}=\delta + e_1/(2zt)$ is taken into account. In
particular the formulae (\ref{C1},\ref{C2}) for the matrices ${\bf C_1}$ and
 ${\bf C_2}$ carry over. But due to the different form of (\ref{Edef_kag})
the matrix ${\bf E}$ is changed compared to (\ref{Eform}) 
\begin{mathletters}
\label{Eform_unfrus}
\begin{eqnarray}
{\bf E} \underline{u} &=& zt/2 \underline{u} + \sqrt{z}t\underline{v}\\
{\bf E} \underline{v} &=& -\sqrt{z}t \underline{u} - t/2\underline{v}
\end{eqnarray}
\begin{equation}
\Rightarrow \widetilde{\bf E} = t\left(
  \begin{array}{cc}
    z/2& -\sqrt{z} \\
    \sqrt{z}& -1/2
  \end{array}
\right)
\end{equation}
\end{mathletters}
acting on $(\underline{u},\underline{v})$. This matrix
is no longer singular as was ${\bf E}$ in (\ref{esimple}). Thus
we stay on the $2\times2$ matrix level. The singularity condition
based on (\ref{geomseries}) is
\begin{equation}
\label{res3_kag}
0=\det\left(1-{\bf C_2}\widetilde{\bf E}^+ {\bf C_1}\widetilde{\bf E} \right)
\end{equation}
which can  be evaluated easily. This concludes the derivation for
the RES3 ansatz on the frustrated kagome lattice.

\section{DOS for the lattices considered} \label{app:dos}
In this appendix we give the explicit formulae for the densities of states
for the lattices discussed in sect.\ \ref{sect:resvarlat}. $K[m]$ stands for 
the complete elliptic integral of the first kind (see e.g.\ \cite{abram64}).

\subsection*{Square lattice}
\begin{equation}
\rho_{\Box} (\varepsilon) = (2|t| \pi^{2})^{-1} \cdot K\left[1 - 
\left(\frac{\varepsilon}{4t}\right)^{2} \right]
\end{equation}

\subsection*{Simple cubic lattice}
\begin{mathletters}
\begin{eqnarray} 
\rho_{sc} (\varepsilon) &=& \pi^{-1} \int_{u_{1}}^{u_{2}} 
\frac{du}{\sqrt{1 - u^2}} \cdot \rho_{\Box} (\varepsilon + 2tu),
 \\
u_{1} &=& {\rm max}(-1, -2 - \varepsilon/(2t))
\\
 u_{2} &=& {\rm min}(1,  2- \varepsilon/(2t))
\end{eqnarray}
\end{mathletters}

\subsection*{bcc lattice}
\begin{equation} 
\rho_{bcc} (\varepsilon) = \frac{2}{\pi} 
\int^{4|t|}_{|\varepsilon|/2} 
\frac{du}{\sqrt{4u^2-\varepsilon^2}} \cdot \rho_{\Box}(u),
\end{equation}

\subsection*{Triangular lattice}
\begin{mathletters}
\begin{equation}
\rho_{\triangle} (\varepsilon) = (\sqrt{z_{0}} \, t \pi^2)^{-1} \cdot
K\left[z_1/z_0\right]
\end{equation}
For $t>0$, $z_{0}$ and $z_{1}$ are given by
\begin{equation} \label{def:z0}
z_{0} = \left\{ \begin{array}{ll}
3 + 2\sqrt{3 - \varepsilon/t} - (\varepsilon/(2t))^{2} & {\rm{for}}\ 
2t \leq \varepsilon \leq 3t
\\
4 \sqrt{3 - \varepsilon/t} & {\rm{for}}\ -6t \leq \varepsilon 
\leq 2t \end{array} 
\right. \ ,
\end{equation}
\ \vspace{-5mm}
\begin{equation} \label{def:z1}
z_{1} = \left\{ \begin{array}{ll}
4 \sqrt{3 - \varepsilon/t}  & {\rm{for}}\ 
2t \leq \varepsilon \leq 3t \\
3 + 2\sqrt{3 - \varepsilon/t} - (\varepsilon/(2t))^{2} & {\rm{for}}
\ -6t \leq \varepsilon \leq 2t \end{array} 
\right. \ .
\end{equation}
\end{mathletters}
For $t<0$, the upper and lower intervals in (\ref{def:z0}) and
(\ref{def:z1}) have to be replaced by
$-3|t| \leq \varepsilon \leq -2|t|$ and  $-2|t| \leq \varepsilon 
\leq 6|t|$, respectively.

\subsection*{Honeycomb lattice}
\begin{equation}
\rho_{\rm H} (\varepsilon) = |\varepsilon/t| \cdot \rho_{\triangle}
(3t - \varepsilon^2/t)
\end{equation}

\subsection*{Kagome lattice}
\begin{equation}
\rho_{\rm K} (\varepsilon) = \frac{1}{3} \, \delta(\varepsilon - 2t) + 
\frac{2}{3} \, |1+\varepsilon/t| \cdot \rho_{\triangle} (3t -
(\varepsilon + t)^2/t)
\end{equation}

\subsection*{Hcp lattice ($t=-1$)}
\begin{mathletters}
\begin{equation} 
\rho_{hcp} (\varepsilon) = \frac{2}{\pi} \int_{0}^{1} dy \, \Xi(y),
\end{equation}
with the integrand
\begin{eqnarray}\nonumber
&&\Xi(y) =\\
&& \left\{ \begin{array}{ll}
\sqrt{-2-\varepsilon_{-}} \, \frac{\rho_{\triangle} (\varepsilon_{-}- 
(2+\varepsilon_{-})y^{2})}{\sqrt{\varepsilon_{+}-\varepsilon_{-} +
(2+\varepsilon_{-})y^2}} &
\\
\quad + \sqrt{2 + \varepsilon_{+}}
\, \frac{\rho_{\triangle} (\varepsilon_{+} - (2 + \varepsilon_{+}) 
y^2)}{\varepsilon_{+} - (2+\varepsilon_{+})y^2 -\varepsilon_{-}} & {\rm{for}}\ 
\varepsilon \leq 0  \\
\sqrt{6 - \varepsilon_{-}} \, \frac{\rho_{\triangle} (\varepsilon_{-} +
(6-\varepsilon_{-})y^2)}{\sqrt{\varepsilon_{+} - \varepsilon_{-} -
(6-\varepsilon_{-})y^2}} & {\rm{for}}\ \varepsilon \geq 0  \end{array}
\right. \ ,
\end{eqnarray}
and
\begin{equation}
\varepsilon_{\pm} = \varepsilon + 2 \left( 1 \pm \sqrt{\varepsilon+4} \right)
\ .
\end{equation}
\end{mathletters}


\end{document}